\documentclass[manuscript]{aastex63}

\submitjournal{The Astrophysical Journal}

\shorttitle{Free energy sources in current sheets}
\shortauthors{Jain et al.}

\begin{document}

\title{Free energy sources in current sheets formed in collisionless plasma turbulence}

\correspondingauthor{Neeraj Jain}
\email{neeraj.jain@tu-berlin.de}

\author{Neeraj Jain}
\affiliation{Zentrum f\"ur Astronomie und Astrophysik, Technische Universit\"at Berlin, Hardenbergstr. 36, D-10623, Berlin, Germany}

\author{J\"org B\"uchner}
\affiliation{Zentrum f\"ur Astronomie und Astrophysik, Technische Universit\"at Berlin, Hardenbergstr. 36, D-10623, Berlin, Germany}


\author{Horia Comi\c sel}
\affiliation{Institute for Space Sciences, P.O. Box MG-23, Atomistilor 409, 077125 Bucharest-Magurele, Romania}

\author{Uwe Motschmann}
\affiliation{Institut f\"ur Theoretische Physik, Technischen Universit\"at Braunschweig Mendelssohnstr. 3, D-38106 Braunschweig, Germany}





\begin{abstract}
Collisionless dissipation of macroscopic energy into heat is an unsolved problem of space and astrophysical plasmas, e.g., solar wind and Earth's magnetosheath. The most viable process under consideration is the turbulent-cascade of macroscopic energy to kinetic-scales where collisionless-plasma-processes dissipate the energy. Space observations and numerical simulations show the formation of kinetic scale current sheets in turbulent plasmas. Instabilities in these CS can provide collisionless dissipation and influence the turbulence. Spatial gradients of physical quantities and non-Maxwellian velocity distribution functions provide the free-energy-sources for CS plasma instabilities. To determine the free-energy-sources provided by the spatial gradients of plasma density and electron/ion bulk velocities in CS formed in collisionless turbulent plasmas with an  external magnetic field $\mathbf{B}_0$, we carried out two-dimensional PIC-hybrid simulations and interpret the results within the limitations of the simulation model.

  We found that  ion-scale CS in a collisionless turbulent plasma are formed primarily by electron shear flows, i.e., electron bulk velocity inside CS is much larger than ion bulk velocity        while the density variations through the CS are relatively small  ($<$ 10\%).   The electron-bulk-velocity and, thus, the current density inside the sheets are directed mainly parallel to $\mathbf{B}_0$. The shear in the perpendicular electron- and ion-bulk-velocities generates  parallel electron- and ion-flow-vorticities. Inside CS, parallel electron-flow-vorticity exceeds the parallel ion-flow-vorticity, changes sign around the CS centers and peaks near the CS edges. An ion temperature anisotropy develops near CS during the CS formation.  It has positive correlation  with the parallel ion- and electron-flow-vorticities. Theoretical estimates support the simulation results.
\end{abstract}



\section{Introduction \label{sec:introduction}}
In collisionless plasmas ranging from
hot laboratory to dilute astrophysical plasmas, irreversible
dissipation of  macroscopic  energy  into heat, without the normal
channels of viscosity and electrical resistivity, is a key unsolved
problem. Turbulent transfer of the energy from macroscopic to micro scales (kinetic
scales such as Larmor radii and inertial lengths of plasma particles), where it is finally dissipated into heat by kinetic plasma
processes, is considered one of the most viable mechanism of the dissipation
in collisionless plasmas \citep{marsch2006}. The energy transfer from macroscopic to kinetic scales takes place by
an anisotropic cascade process mediated by Alfv\'en waves
 \citep{kraichnan1965,goldreich1995,shebalin1983,montgomery1995,zank1992,zank1993,matthaeus1990,bieber1996,howes2015,loureiro2017}.
       At kinetic scales, plasma processes  transfer the energy from turbulent electromagnetic fields to kinetic energy of plasma particles by  field-particle interactions and to internal energy of plasma by pressure-strain interaction \citep{yang2017a,yang2017b,chasapis2018}. The irreversible dissipation is supposed to be realized by an entropy cascade in velocity space to scales small enough that  even the infrequent collisions can thermalize a fair number of particles \citep{schekochihin2009}.

 At kinetic scales, both the wave-particle resonance processes distributed in the volume of plasma \citep{hollweg2002,leamon1998a,bale2005,schekochihin2009,howes2011,podesta2012,gary2016,chandran2010,bourouaine2013} and plasma processes localized in coherent structure \citep{greco2012,servidio2012,dmitruk2004,sundkvist2007,parashar2016,chasapis2017,karimabadi2013,wan2015,osman2011,osman2014} dissipate the turbulence energy.
An increasing number of observational and simulation studies in
recent years support the intermittent dissipation
localized in and around kinetic scale current sheets which form self 
consistently in plasma 
turbulence and contain significant power of the turbulence \citep{matthaeus2015,borovsky2010}.
Current sheets with thicknesses ranging from ion to electron scales are
observed ubiquitously in numerical simulations and space observations
of collisionless plasma turbulence \citep{sundkvist2007,biskamp1989,maron2001,franci2015,perri2012,howes2016,podesta2017}. 
Therefore, an understanding of kinetic plasma processes
 in current sheets formed in kinetic plasma turbulence is crucial to
 solve the puzzle of dissipation and heating in turbulent
 collisionless plasmas, e.g., the solar wind, solar corona and Earth's
 magnetosphere.

  Several plasma processes, mainly, stochastic ion heating \citep{chen2001,chandran2010,markovskii2011}, Landau and cyclotron damping \citep{tenbarge2013}, and acceleration by parallel electric fields \citep{hoshino2001,drake2005,fu2006,oka2010a,egedal2012} and  Fermi acceleration in contracting magnetic islands \citep{drake2006,oka2010b,dahlin2014,li2015,zank2014,leroux2015} generated by magnetic reconnection, have been proposed as collisionless  mechanisms of the energy dissipation in current sheets. The role of magnetic reconnection in turbulence is being discussed since it was first proposed  in  Nineteen-eighties \citep{matthaeus1980,matthaeus1984,matthaeus1986,ambrosiano1988,lazarian1999}. Fully kinetic three dimensional simulations allowed by the increase in the computational power in the later years strongly suggest the role of reconnection driven turbulence in plasma heating and particle acceleration \citep{daughton2011,dahlin2015,li2019}. Stochastic ion heating in the current sheets can occur during their rapid formation in collisionless plasma turbulence \citep{markovskii2011} and/or while current sheets undergoing magnetic reconnection \citep{yoon2019}.
  These plasma processes at kinetic scales are either directly or indirectly related to plasma instabilities. Magnetic reconnection in current sheets formed in plasma turbulence is a time-dependent tearing/plasmoid instability like process  (which allow breaking of frozen in condition of magnetic field and reconnection of magnetic field lines in current sheets) and can be influenced by several other plasma instabilities in current sheet \citep{ergun_egu2017,munoz2017}.
Tearing instability of an ion-scale current sheet can be
 enhanced by ion cyclotron instability while suppressed by firehose
 instability \citep{gingell2015}.
 Stochastic ion heating is also known to be triggered by plasma instabilities \citep{demchenko1974,hendel1973,stasiewicz2020}.

Plasma instabilities in current sheets have also been suggested to influence the location of the ion scale break observed in the power spectrum of plasma turbulence. Laboratory experiments of magnetic
reconnection strongly suggest a connection between the
ion-scale spectral break, observed near lower hybrid frequency in the
experiments, and instabilities of a single current sheet \citep{stechow2016}.
In solar wind turbulence, the location (near ion cyclotron frequency) of the ion-scale break  was linked to the current sheet thickness
which
is a crucial parameter to determine the growth of the plasma instabilities in current sheets \citep{podesta2015}.
It was also observed to depend on the 
amplitude of the magnetic field fluctuations \citep{markovskii2008} 
which can be controlled by plasma instabilities in current sheets.

Numerical simulations, space observations, laboratory experiments and theoretical studies  of collisionless plasma
turbulence and individual current sheets, therefore, suggest that plasma
instabilities in current sheets formed in plasma turbulence  can play an important role not only in collisionless dissipation but also in the kinetic scale properties of the observed
turbulence spectra, in particular the spectral breaks.
Growth of plasma instabilities depends on free energy sources available from spatial gradients  of physical quantities and/or from non-Maxwellian features of plasma particles' distribution functions. A clear understanding of the free energy sources available in the current sheets formed in collisionless plasma turbulence is, therefore,  essential to pin  point the role of plasma instabilities in the turbulence.  

In a current sheet, current density $\mathbf{J}=ne(\mathbf{u}_i-\mathbf{u}_e)$ ($n$ and $\mathbf{u}_{i,e}$  are plasma number density and ion/electron bulk velocities, respectively) is confined in a small thickness. Therefore, free energy sources in a current sheet can come from spatial gradients of $n$, $\mathbf{u}_{i}$ and $\mathbf{u}_{e}$. In addition, the non-Maxwellian features in the forms of temperature anisotropy and/or relative drift of plasma particles might also be present. Ion temperature anisotropy is often observed near the current sheets formed in collisionless plasma turbulence. Relative contributions of $n$, $\mathbf{u}_i$ and $\mathbf{u}_e$ in the formation of the current sheets in the turbulence are, however, not known yet.

In this paper, we carry out 2-D PIC-hybrid simulations of collisionless plasma turbulence to study the relative contributions of $n$, $\mathbf{u}_i$ and $\mathbf{u}_e$ in the formation of the current sheets in the turbulence. We find that ion-scale current sheets are formed primarily by electron shear flow, i.e., electron bulk velocity is much larger than ion bulk velocity and density variation is relatively small ($<$ 10\%) inside current sheets. Electron bulk velocity and thus current inside sheets are directed mainly parallel to the external magnetic field. Shear flow in perpendicular bulk velocities of electrons and ions generates parallel components of electrons and ions vorticity, the former of which is larger than the later  inside current sheets, changes sign around the center and peaks near the edges of current sheets. Ion temperature anisotropy develops near current sheets during the formation of current sheets and  has positive correlation  with both the electron and ion vorticities. Theoretical estimates in the limit of un-magnetized ions support the simulation results.

The paper is organized as follows. Section \ref{sec:simulation} describes the simulation setup. Results are presented in section \ref{sec:results}. Theoretical estimates are presented in section \ref{sec:theory}. Discussion of  results and conclusion are presented in  section \ref{sec:conclusion}.

\section{Simulation setup\label{sec:simulation}}
We  employ a
hybrid model of plasmas in which ions are treated as particles while electrons as an inertia-less
fluid. Such plasma model leaves out electron inertial and electron kinetic effects important at
electron scales. Our 2-D simulations are carried out using a PIC-hybrid code A.I.K.E.F.  of the
Technical University Braunschweig \citep{mueller2011}. We initialize our 2-D simulations in an x-y plane with random phased fluctuations of magnetic field and plasma velocity imposed on an isotropic background plasma of uniform density $n_0$. A uniform magnetic field $B_0\hat{z}$ is applied perpendicular to the simulation plane. Magnetic field fluctuations are calculated from magnetic vector potential
\begin{eqnarray}
  \tilde{\mathbf{A}}&=&\hat{z}\sum_{k_x,k_y}\delta A_z(k_x,k_y) \sin(k_xx+k_yy+\phi(k_x,k_y)), 
  \end{eqnarray}
where $k_x$ and $k_y$ are wave numbers in x- and y-direction, respectively, and $\phi$ is the wave-number dependent  random phase. The amplitude $\delta A_z$ of the magnetic vector potential is so chosen that the amplitude of magnetic field fluctuation $\delta B_{\perp}=\delta A_z k_{\perp}$ is independent of the wave number, i.e., all initialized modes have the same energy. Plasma velocity fluctuations have the same form as magnetic field fluctuations except the random phases so that the magnetic and velocity fluctuations have vanishing correlation but equi-partition of  energy.   

We initialize fluctuations in the wave number range $|k_{x,y}d_i| < 0.2$ ($k_{x,y}\neq 0$) to have a root-mean-square value $B_{rms}/B_0=0.24$. Here $d_i=v_{Ai}/\omega_{ci}$, $v_{Ai}=B_0/\sqrt{\mu_0n_0m_i}$ and  $\omega_{ci}=eB_0/m_i$  are inertial length, Alfv\'en velocity and cyclotron frequency of ions, respectively, and, $\mu_0$ (vacuum magnetic  permeability), $e$ (electronic charge) and $m_i$ (proton mass) are physical constants.   Electron and ion plasma beta are $\beta_e=2\mu_0n_0k_BT_e/B_0^2=0.5$ and $\beta_i=2\mu_0n_0k_BT_i/B_0^2=0.5$, respectively. Here $T_e$ and $T_i$  are electron and ion temperatures, respectively, and $k_B$ is the Botzmann constant. The simulation box size $256 d_i \times 256 d_i$ is fixed for $512 \times 512$, $1024 \times 1024$ and $2048 \times 2048$ grid points with 500, 1000 and 2000 particles per cell, respectively. The time step for the three grid sizes are $\Delta t$=0.01, 0.0025 and 0.001 $\omega_{ci}^{-1}$, respectively. Such small values of time step allows us to take collisional resistivity as zero in all simulations. Boundary conditions are periodic in all directions.

\begin{figure}
  \begin{center}
  \includegraphics[clip=true,trim=0cm 0cm 0cm 0cm,width=0.49\linewidth]{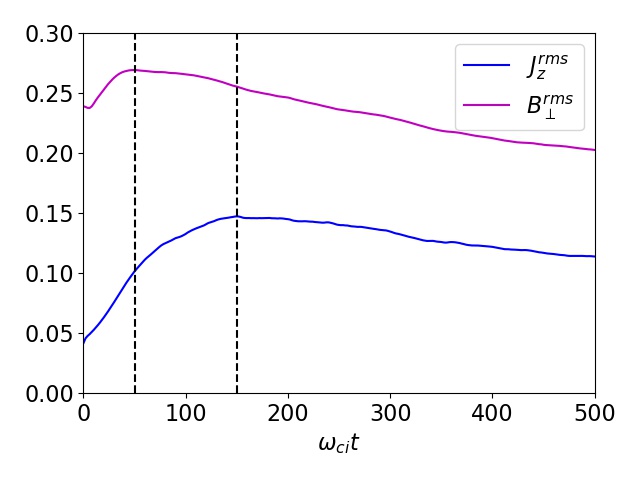}\\
    \includegraphics[clip=true,trim=0cm 0cm 0cm 0cm,width=0.49\linewidth]{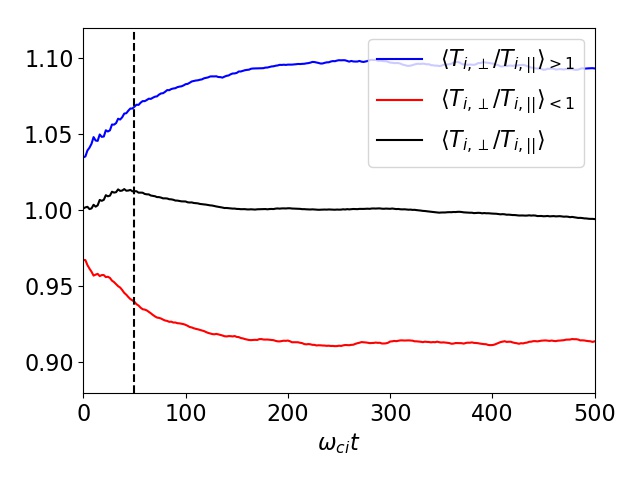}  
    \caption{Evolution of root-mean-square values of perpendicular magnetic field $B_{\perp}^{rms}/B_0$
      and parallel current density $J_z^{rms}/(n_0 e v_{Ai})$ (top panel). Evolution of ion temperature anisotropy $T_{i,\perp}/T_{i,||}$ averaged over the grid locations where $T_{i,\perp}/T_{i,||} > 1$ (perpendicular temperature anisotropy $\langle T_{i,\perp}/T_{i,||}\rangle_{>1}$), the locations where $T_{i,\perp}/T_{i,||} < 1$ (parallel temperature anisotropy $\langle T_{i,\perp}/T_{i,||}\rangle_{<1}$) and all the locations (net temperature anisotropy $\langle T_{i,\perp}/T_{i,||}\rangle$) in the simulation domain    (bottom panel).   Vertical dashed lines are drawn at $\omega_{ci}t=50$ and $\omega_{ci}t=150$ (only in top panel). \label{fig:rms_evolution}}
\end{center}
\end{figure}
\section{Simulation results \label{sec:results}}
\begin{figure}
  \begin{center}
  \includegraphics[clip=true,trim=0.8cm 1.7cm 0.6cm 0cm,width=0.75\linewidth]{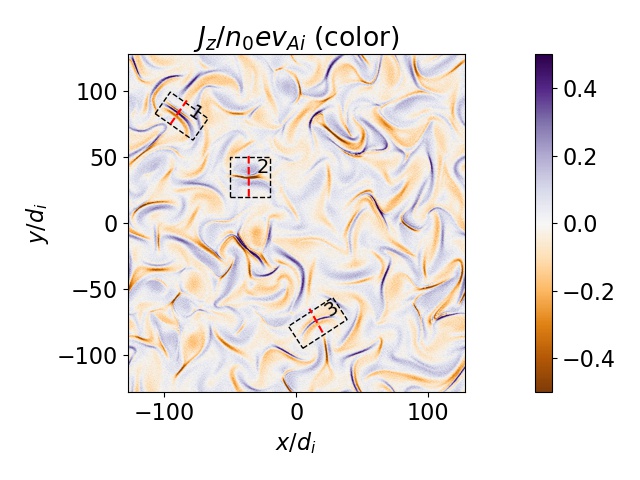}
    \includegraphics[clip=true,trim=0.8cm 1cm 0.6cm 1.85cm,width=0.75\linewidth]{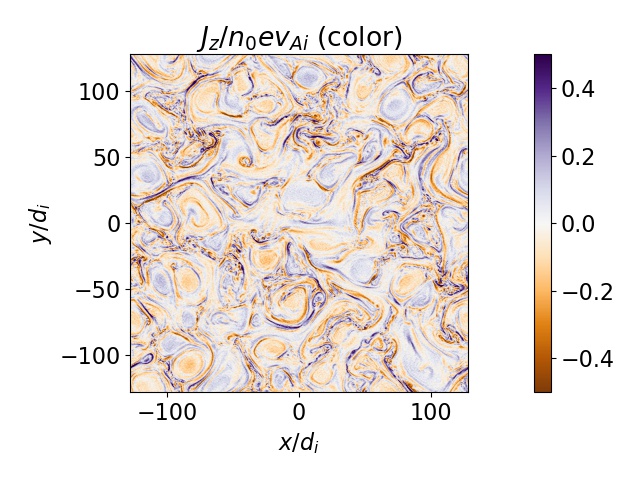}
    \caption{Out-of-plane current density $J_z$ in the x-y simulation plane at $\omega_{ci}t=50$ (top) and 150 (bottom). Three current sheets selected for detailed analysis are highlighted in the top panel by enclosing them  in boxes and are numbered 1-3. \label{fig:jz_t50_t150}}
    \end{center}
  \end{figure}

The random-phased fluctuations of magnetic field and ion bulk velocity initialized at long wavelength in our simulations  evolve to form current sheets.  Evolutions of  root-mean-square (RMS) values of perpendicular magnetic field $\mathbf{B}_{\perp}$ and  parallel current density $J_z$ 
are shown in Fig. \ref{fig:rms_evolution}, where  RMS value of a variable $\psi$ is calculated as
\begin{eqnarray}
  \psi^{rms}=[\langle \psi^2 \rangle -\langle \psi \rangle^2]^{1/2}.
\end{eqnarray}
Here parallel and perpendicular directions are with respect to the (z-) direction of the applied magnetic field. 
RMS values grow from their initial values to reach maximum values and then decay slowly.
The time of reaching maximum for  $B_{\perp}^{rms}$,  $\omega_{ci}t=50$, is different from that for  $J_z^{rms}$ (at $\omega_{ci}t=150$).
Fig. \ref{fig:jz_t50_t150} shows parallel current density $J_z$ in the whole simulation domain at the two times when $B_{\perp}^{rms}$ and $J_z^{rms}$ reach maximum. Current sheets get formed in the turbulence by $\omega_{ci}t=50$. A typical current sheet at $\omega_{ci}t=50$ has  a central current accompanied by return side currents (opposite to the central current) providing current closure. The central current has relatively sharper variation. These current sheets break up developing their own turbulence by $\omega_{ci}t=150$ \citep{daughton2011,munoz2017,dahlin2015}. 

Fig. \ref{fig:rms_evolution} also shows the evolutions of various averages of the ion temperature anisotropy $T_{i,\perp}/T_{i,||}$. Net average ion temperature anisotropy $\langle T_{i,\perp}/T_{i,||}\rangle$, obtained by averaging $T_{i,\perp}/T_{i,||}$ over the whole simulation grid, grows to reach a peak a little before $\omega_{ci}t=50$ (the time by which current sheets have formed) with a value slightly above its initial isotropic value of unity  and then drops to saturate around the isotropic value,  consistent with other hybrid simulations \citep{franci2015}.  On the other hand, average perpendicular  and parallel   ion temperature anisotropies, $\langle T_{i,\perp}/T_{i,||}\rangle_{>1}$ and $\langle T_{i,\perp}/T_{i,||}\rangle_{<1}$ (obtained by averaging $T_{i,\perp}/T_{i,||}$ over the simulation grid locations where  $T_{i,\perp}/T_{i,||}>1$ and $T_{i,\perp}/T_{i,||}<1$, respectively) continue to develop beyond the peak of $\langle T_{i,\perp}/T_{i,||}\rangle$ and saturate later to the values $\approx$ 1.10 and 0.90 ($10\%$ anisotropy), respectively. This means that both the perpendicular and parallel temperature anisotropies exist in turbulence despite the net temperature anisotropy indicating isotropy.

Our interest in this paper is in free energy sources empowering plasma instabilities in current sheets formed in turbulence.
Therefore we examine these current sheets at $\omega_{ci}t=50$ to look for available  free energy sources well before their depletion by the  growth of plasma instabilities. The time $\omega_{ci}t=50$ is also the time of the perpendicular magnetic energy $\int |\mathbf{B}_{\perp}|^2 dx dy \propto (B_{\perp}^{rms})^2$ reaching maximum (Fig. \ref{fig:rms_evolution}).  Current sheets store magnetic energy and therefore the time of magnetic energy reaching maximum can be taken as the time of peak activity of current sheet formation. Analysis of  current sheets at other times  shows that the conclusions presented in this paper are independent of the choice of the analysis time around  $\omega_{ci}t=50$. 

\begin{figure*}
    \includegraphics[clip=true,trim=0.25cm 1.35cm 0.25cm 0.7cm,width=0.49\linewidth]{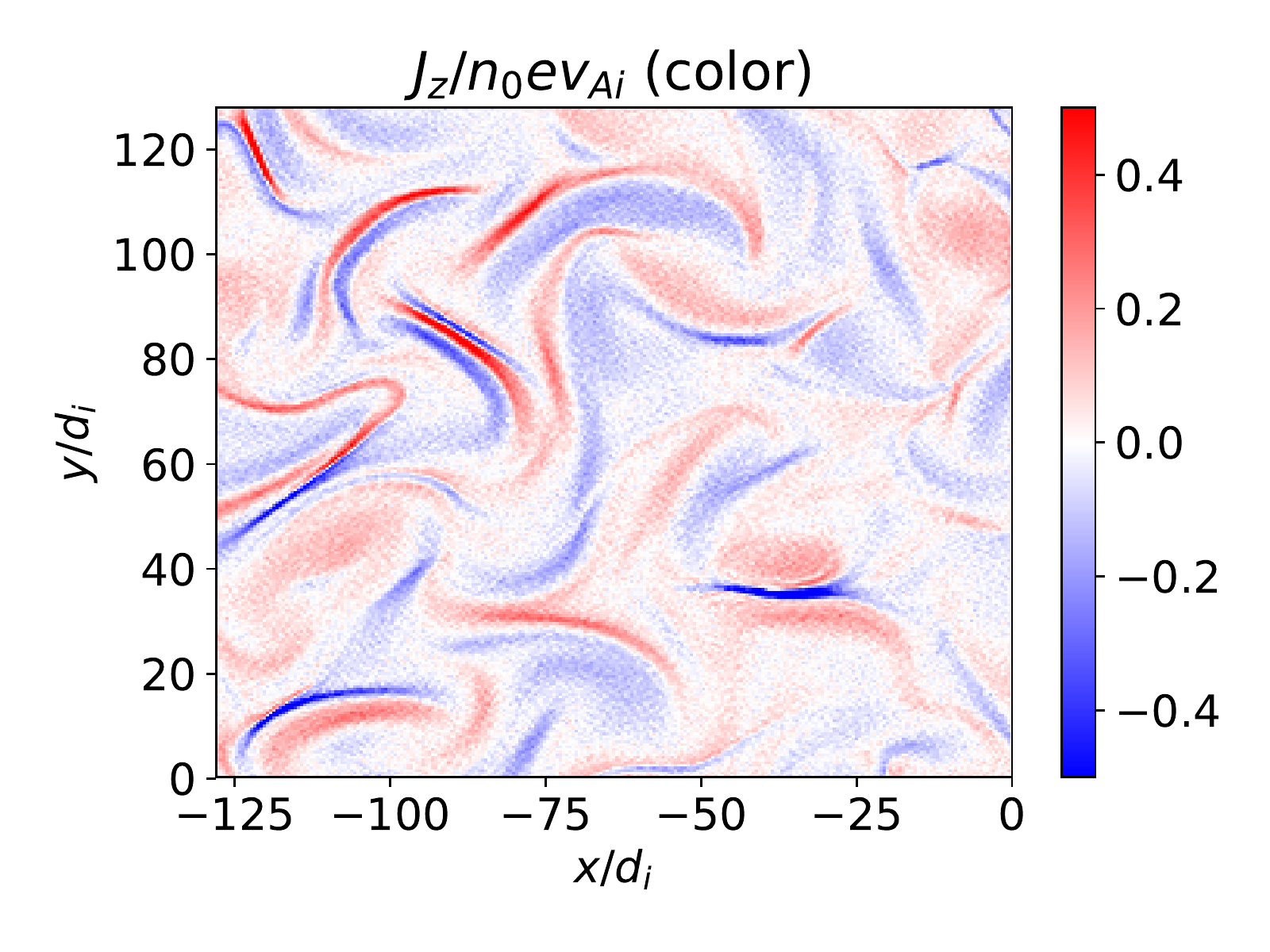}
    \put(-190,140){\large (a)}
        \includegraphics[clip=true,trim=0.25cm 1.35cm 0.25cm 0.7cm,width=0.49\linewidth]{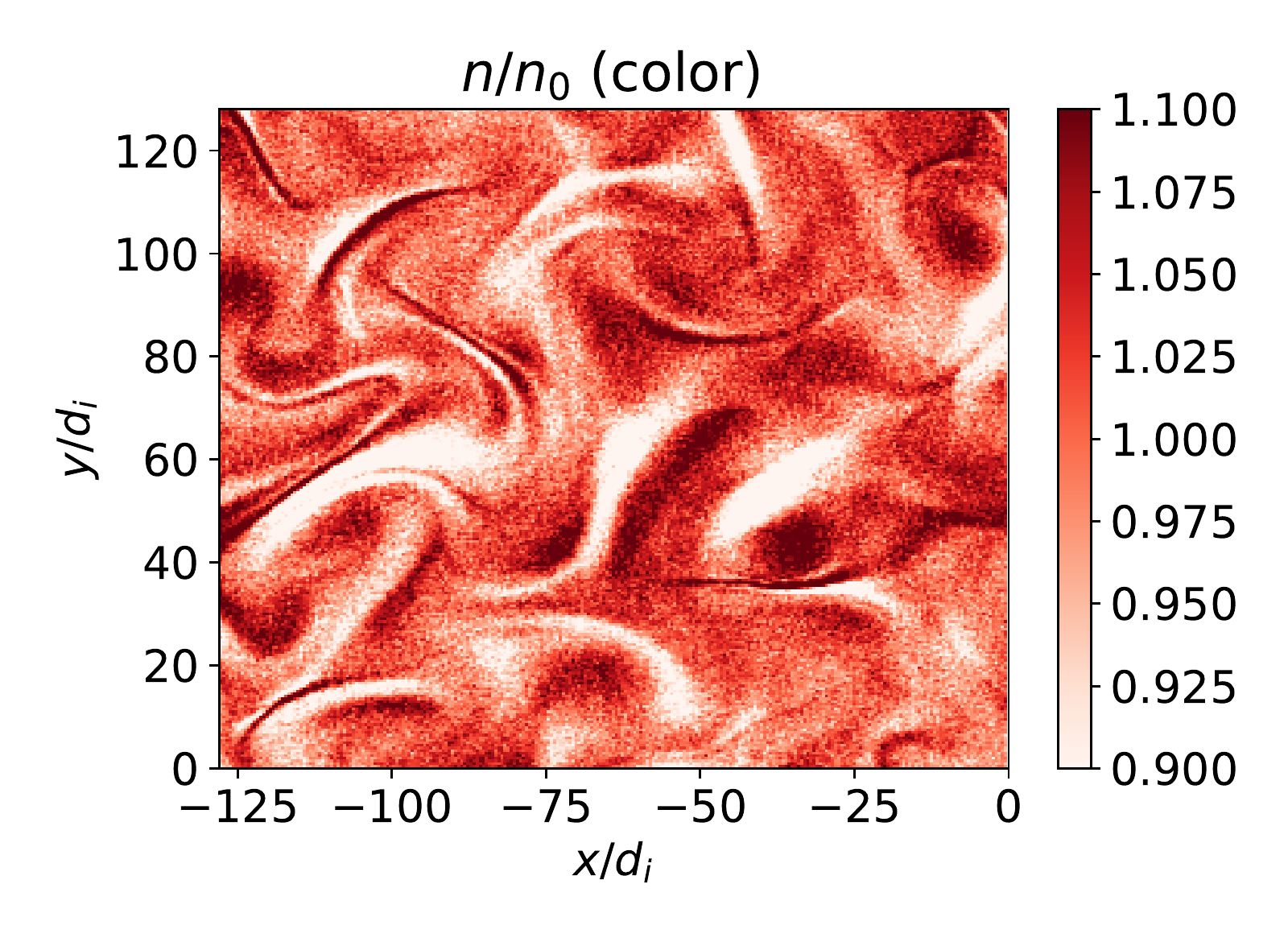}
        \put(-190,135){\large (b)}
    \vspace{0.1in}

    \includegraphics[clip=true,trim=0.25cm 0.7cm 0.25cm 0.7cm,width=0.49\linewidth]{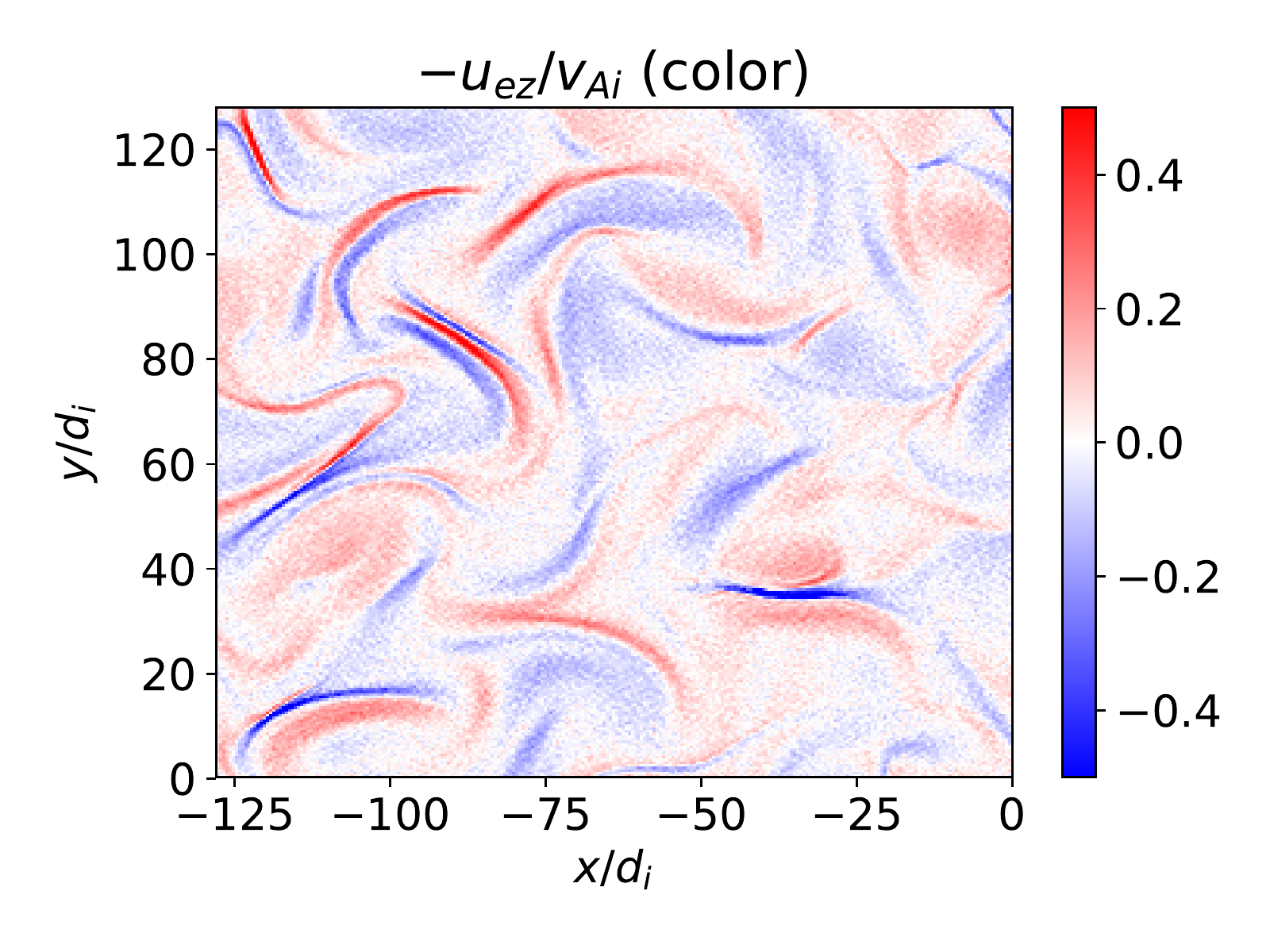}
    \put(-190,150){\large (c)}
    \includegraphics[clip=true,trim=0.25cm 0.7cm 0.25cm 0.7cm,width=0.49\linewidth]{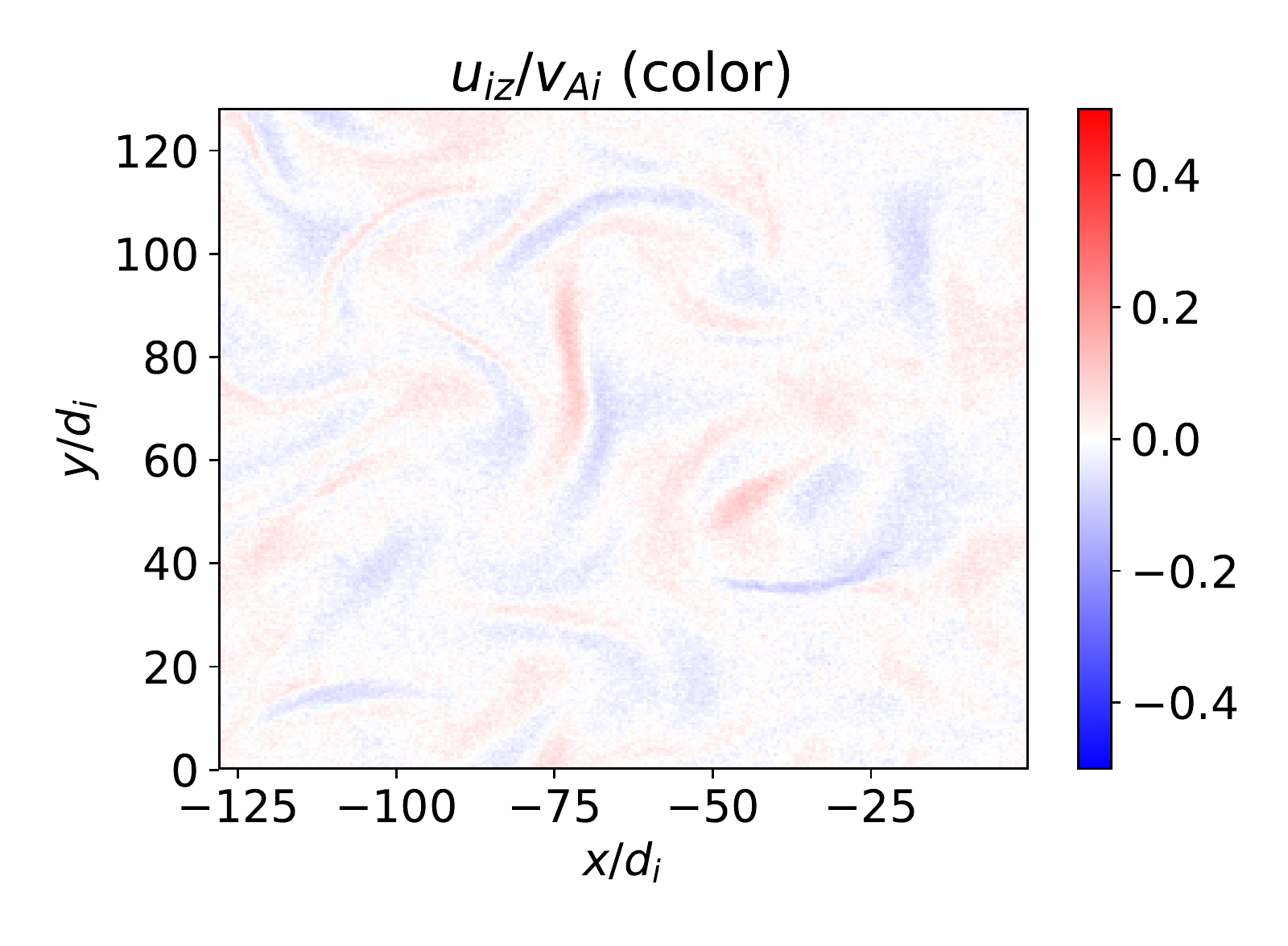}
    \put(-190,145){\large (d)}
      \caption{Parallel current density $J_z$ (a), plasma number density $n$ (b), negative of parallel electron bulk velocity $-u_{ez}$ (c) and  parallel ion bulk velocity (d)  in the top-left quadrant of the simulation domain at $\omega_{ci}t=50$. \label{fig:ue_ui_n_t50}}
\end{figure*}

\begin{figure*}
  \begin{center}
  \includegraphics[clip=true,trim=0.25cm 1.35cm 0.25cm 0.7cm,width=0.45\linewidth]{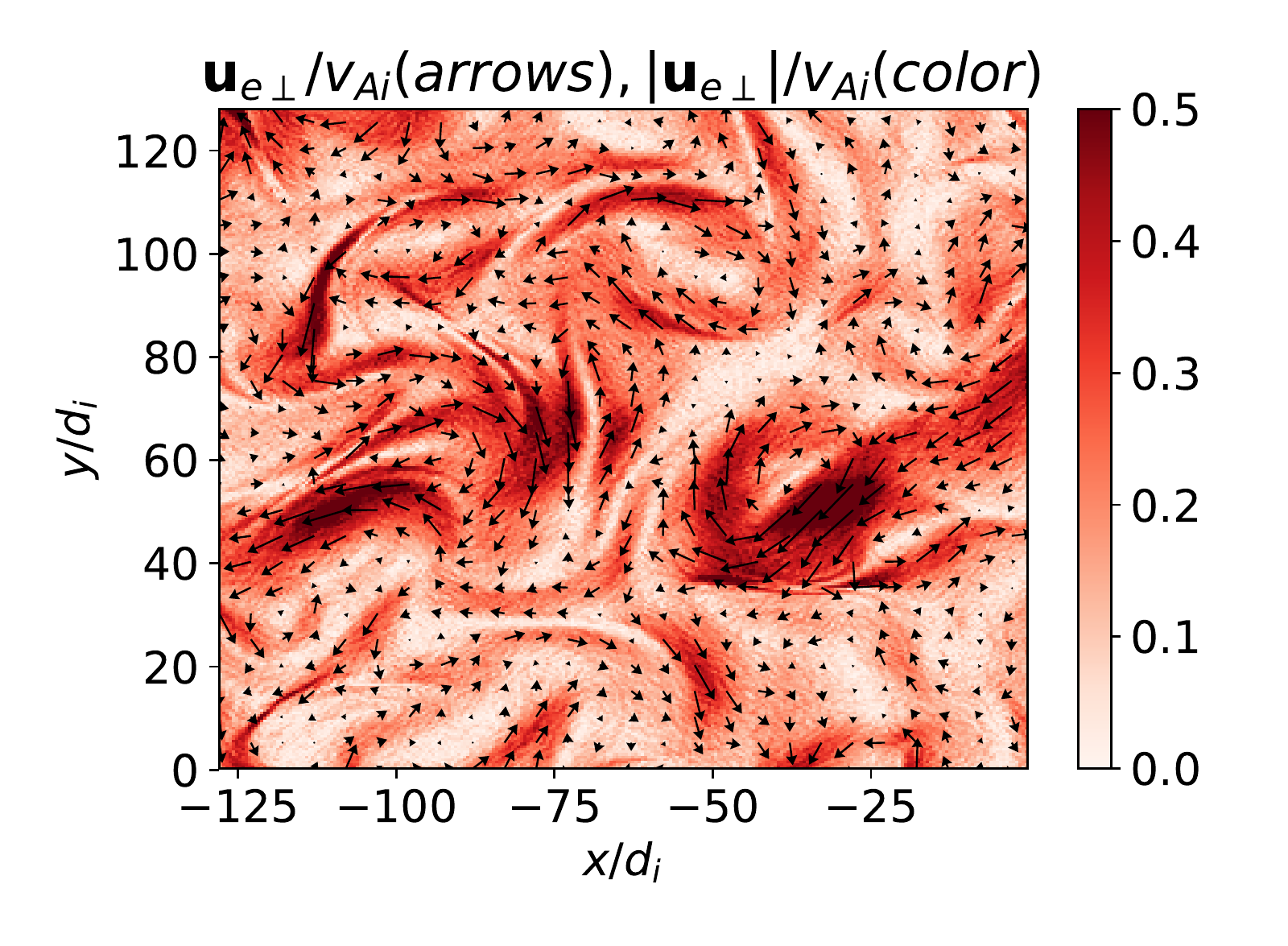}
      \put(-220,140){\large (a)}
  \includegraphics[clip=true,trim=0.25cm 1.35cm 0.25cm 0.7cm,width=0.45\linewidth]{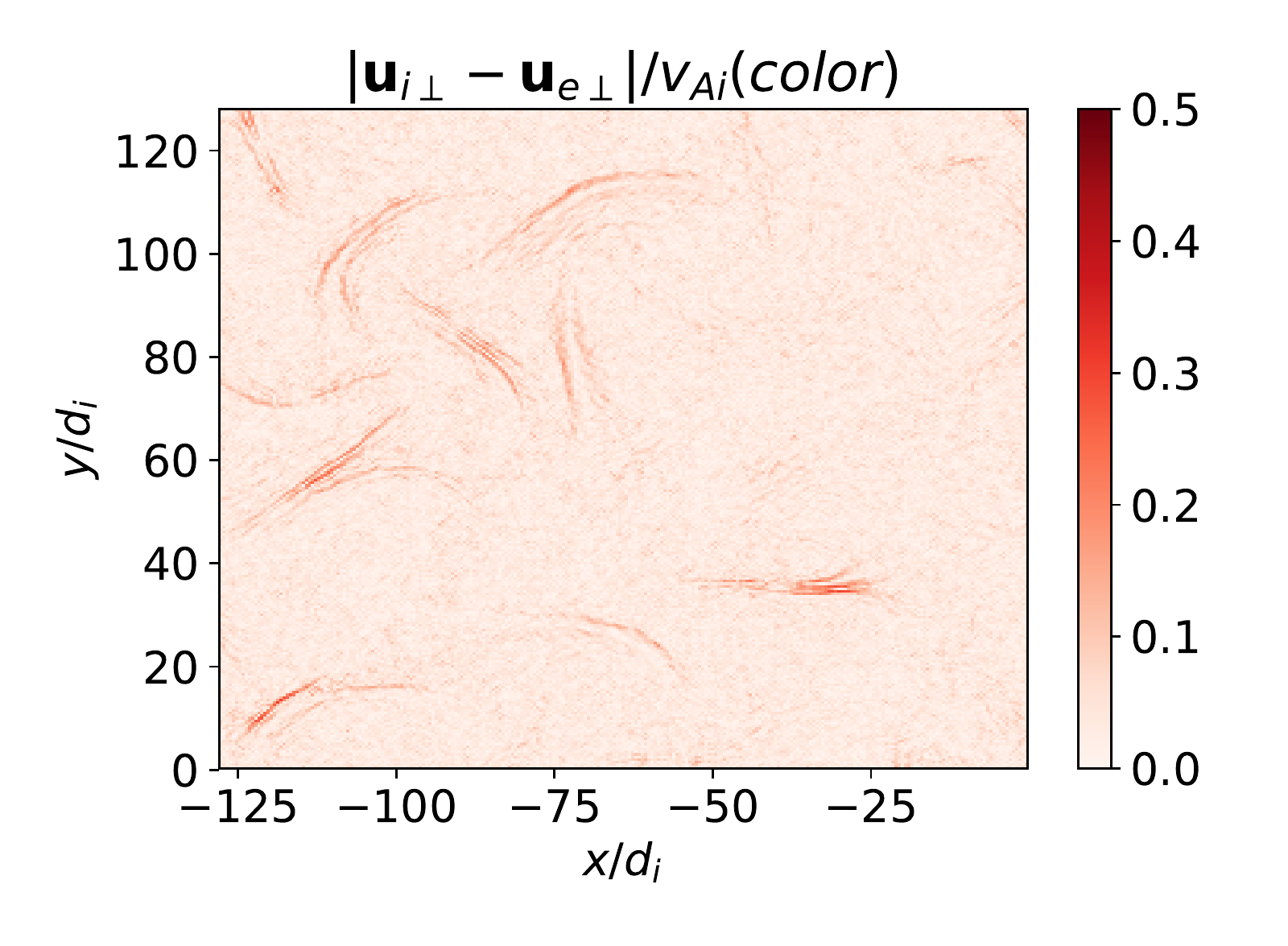}
  \put(-220,140){\large (b)}
  \vspace{0.1in}
  
  \includegraphics[clip=true,trim=0.25cm 0.7cm 0.25cm 0.7cm,width=0.45\linewidth]{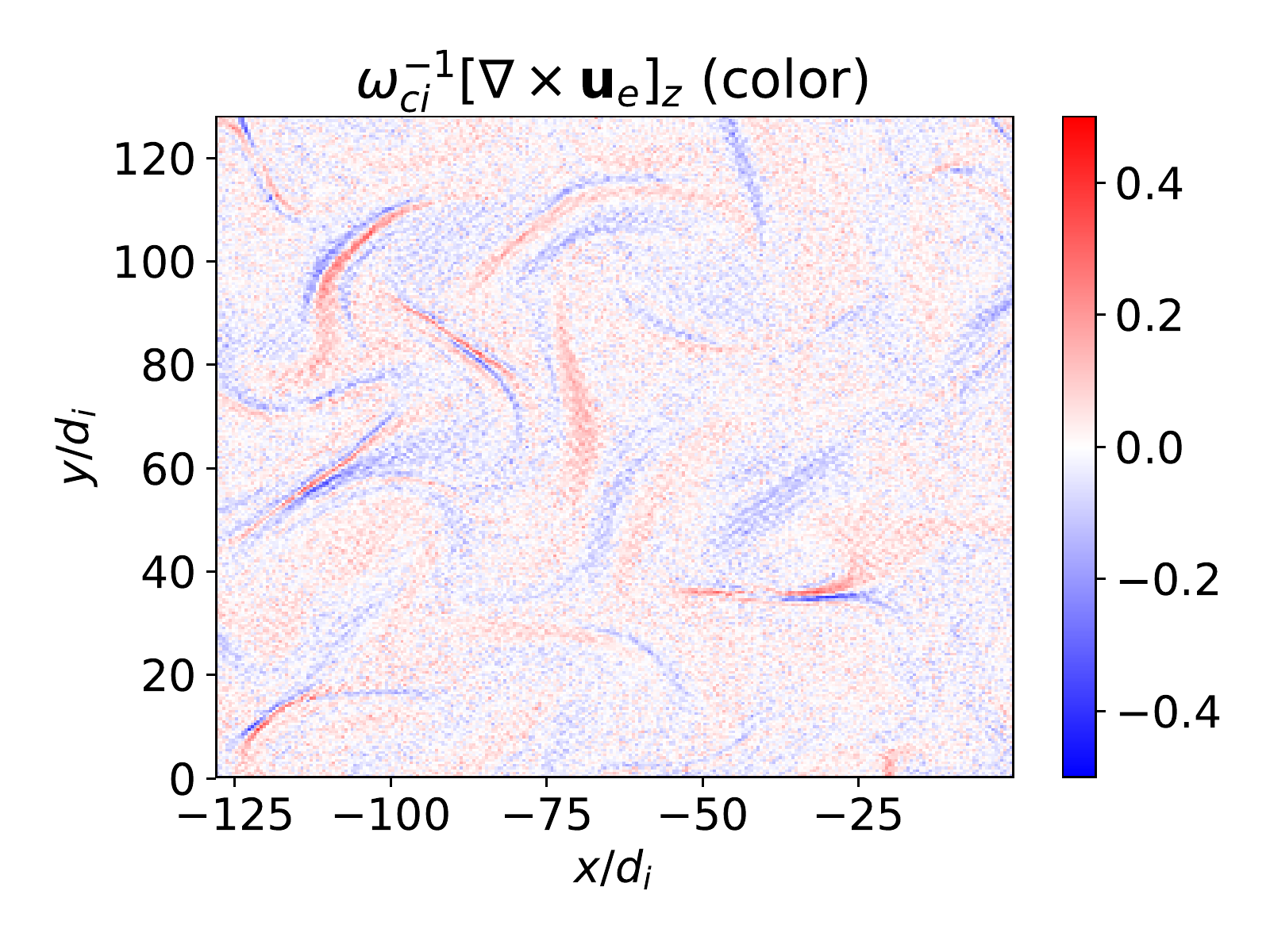}
      \put(-215,150){\large (c)}
      \includegraphics[clip=true,trim=0.25cm 0.7cm 0.25cm 0.7cm,width=0.45\linewidth]{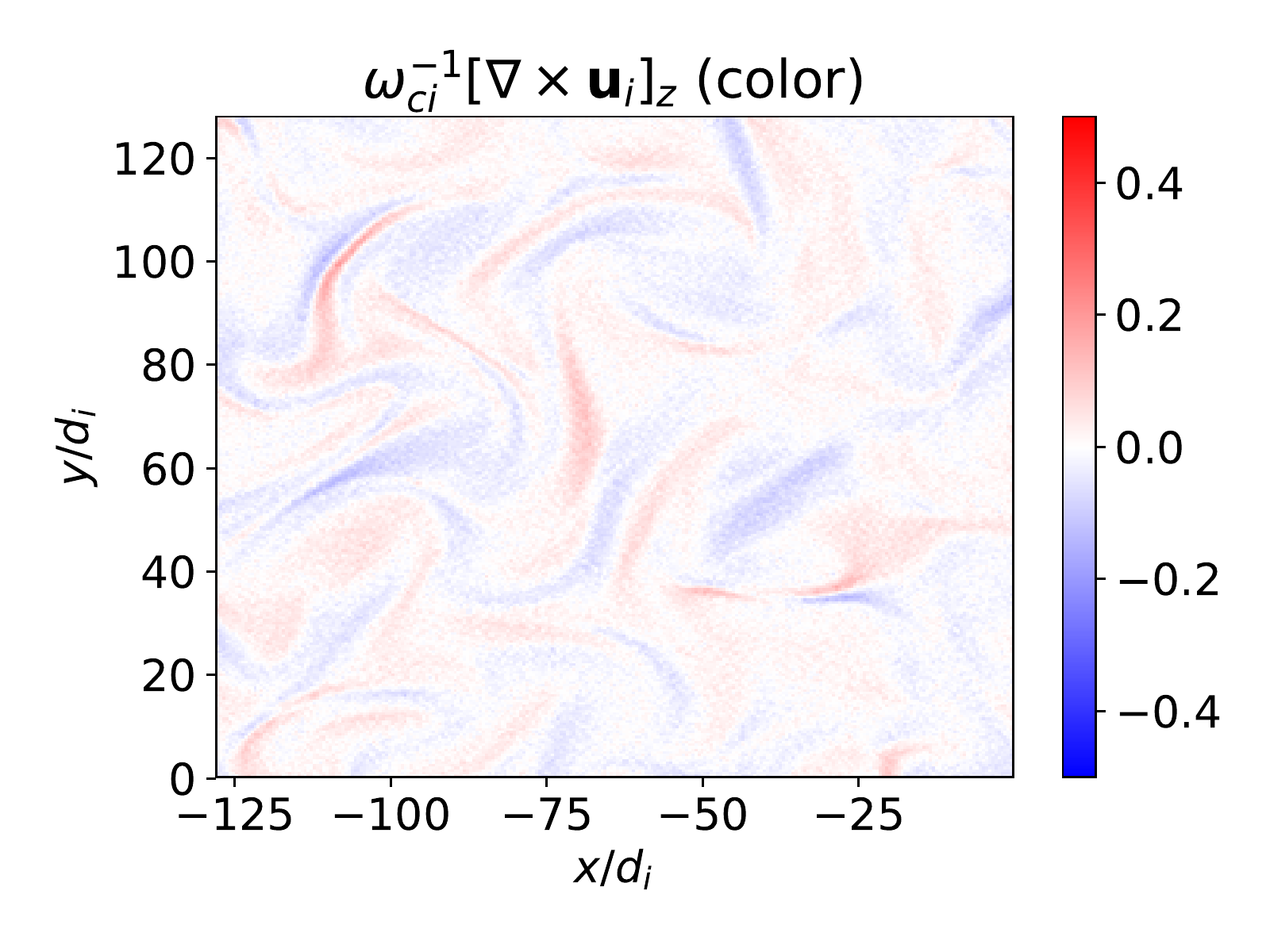}
          \put(-215,150){\large (d)}
          \caption{Magnitude (color) and vectors (arrows) of perpendicular electron bulk velocity (a), magnitude of the difference of perpendicular ion and electron  bulk velocities (b), parallel electron (c) and ion (d) vorticities shown in  the top-left quadrant of the full simulation domain at $\omega_{ci}t=50$. \label{fig:ueperp_uiperp_t50}}
          \end{center}
\end{figure*}

\begin{figure*}
  \begin{center}
\includegraphics[width=0.32\linewidth]{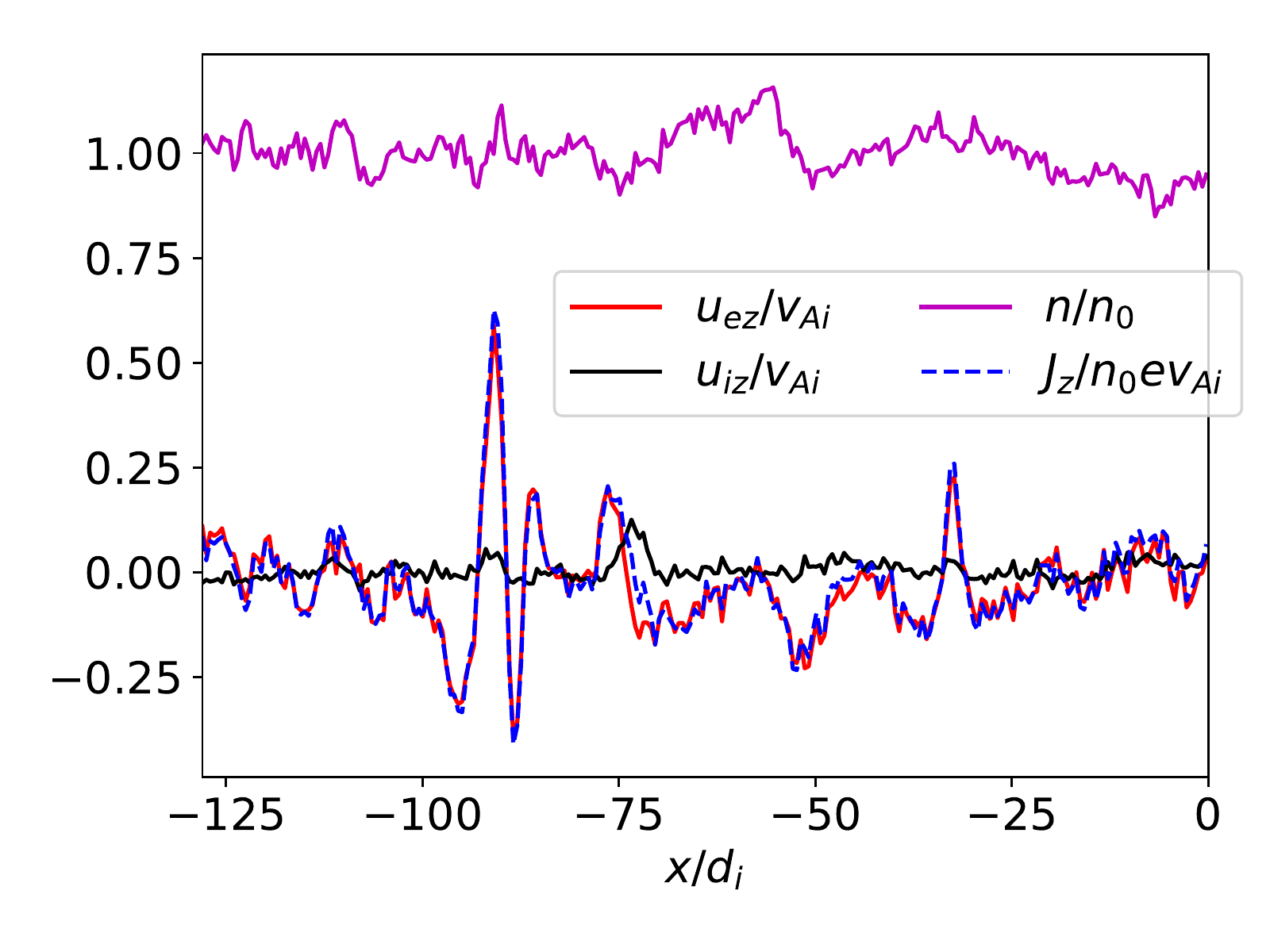}
\put(-75,120){\large (a)}    
\includegraphics[width=0.32\linewidth]{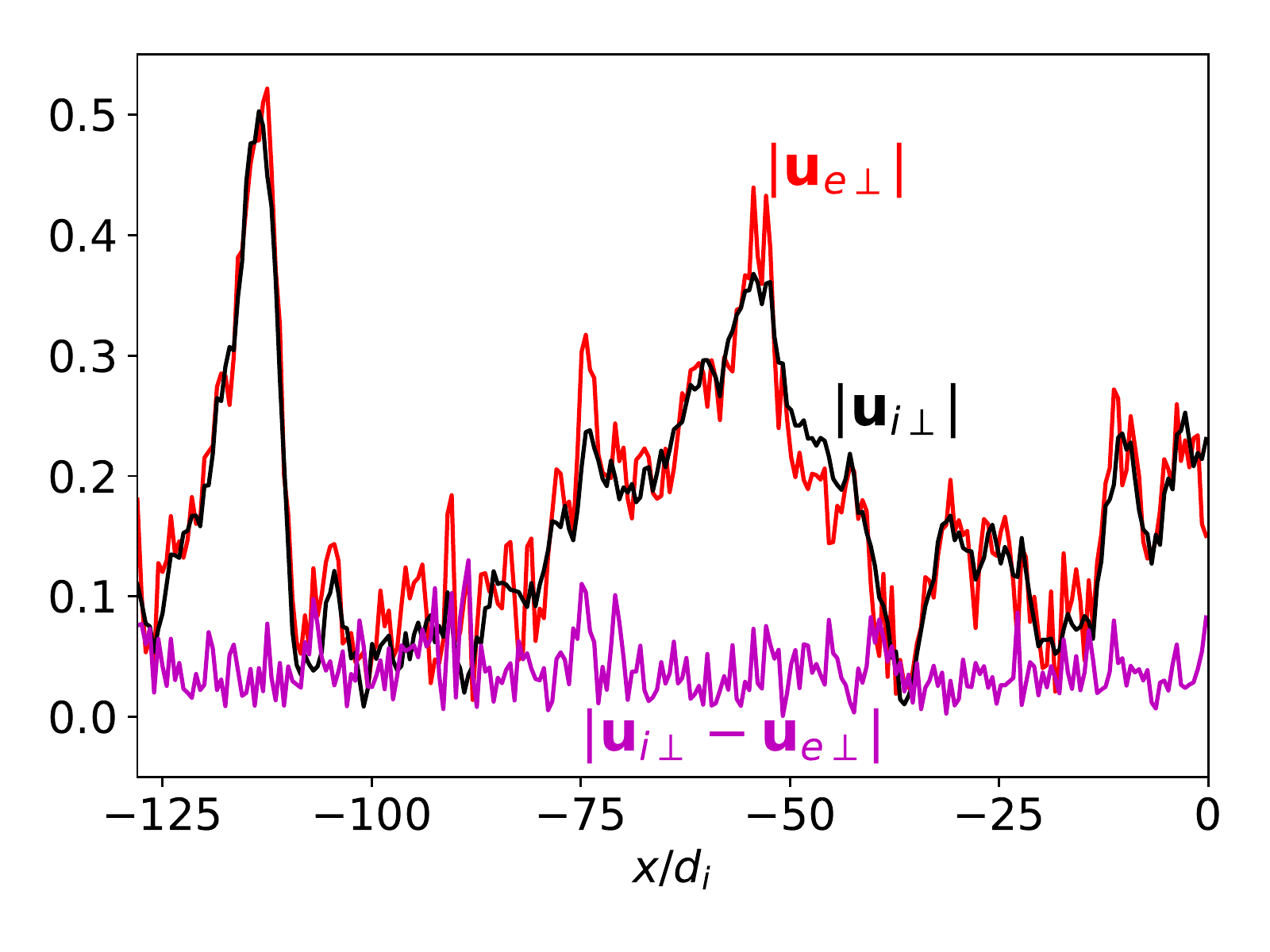}
\put(-75,120){\large (b)}    
\includegraphics[width=0.32\linewidth]{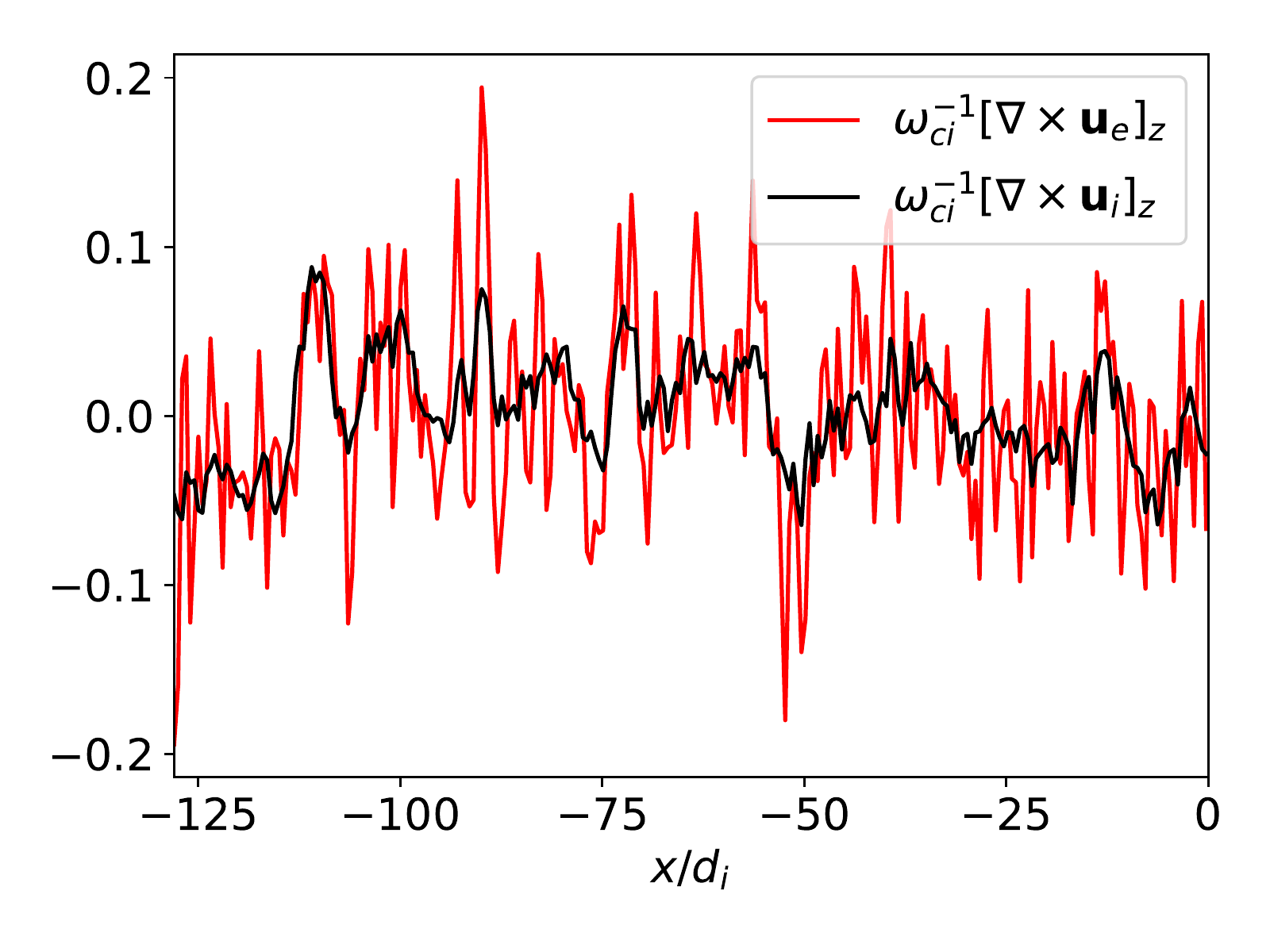}
\put(-75,120){\large (c)}    
  \caption{Line-outs along the line $y/d_i\approx 85$ at $\omega_{ci}t=50$. (a) $n$, $u_{ez}$, $u_{iz}$ and $J_z$; (b) $|\mathbf{u}_{e\perp}|$, $|\mathbf{u}_{i\perp}|$ and $|\mathbf{u}_{i\perp}-\mathbf{u}_{e\perp}|$; (c) $[\nabla \times \mathbf{u_e}]_z$ and $[\nabla \times \mathbf{u}_i]_z$. \label{fig:xlineouts}}
\end{center}
  \end{figure*}

Figs. \ref{fig:ue_ui_n_t50} and \ref{fig:ueperp_uiperp_t50} show various quantities at $\omega_{ci}t=50$ in a quarter  of the simulation plane (the top-left quadrant of the planes shown in Fig. \ref{fig:jz_t50_t150}) to inspect the free energy sources provided by the  spatial gradients of $n$, $\mathbf{u}_e$ and $\mathbf{u}_i$ in current sheets. Fig. \ref{fig:xlineouts} shows the line-outs of these quantities along a randomly chosen line $y/d_i$=85.
It is evident  that parallel current density $J_z$ in current sheets (Fig. \ref{fig:ue_ui_n_t50}a) is almost entirely contributed by parallel electron bulk velocity $u_{ez}$ (Fig. \ref{fig:ue_ui_n_t50}c)     which is much larger than the parallel ion bulk velocity $u_{iz}$ inside current sheets (Fig. \ref{fig:ue_ui_n_t50}d). Line-outs in Fig. \ref{fig:xlineouts}a show that $|u_{iz}| < |u_{ez}|$ except when $J_z$ is very small.
Plasma number density has strong gradients inside current sheets in comparison to out-side current sheets but its variation (under 10\% about the mean value $n/n_0=1$) inside current sheets does not affect significantly the current sheet structure.   Changes in density are due to the  slower ion dynamics and expected to be smaller compared to the changes in electron bulk velocity made by much faster electron dynamics.  This has been seen in PIC simulations of collisionless guide field magnetic reconnection (presented in the appendix of \cite{jain2017}) which show that the spatial variation of ion and electron densities in current sheet (where parallel bulk velocity of electrons is much larger than that of ions) is at most 10\%.

\begin{figure}
  \begin{center}
    \includegraphics[clip=true,trim=0.25cm 0.7cm 0.25cm 0.7cm,width=0.49\linewidth]{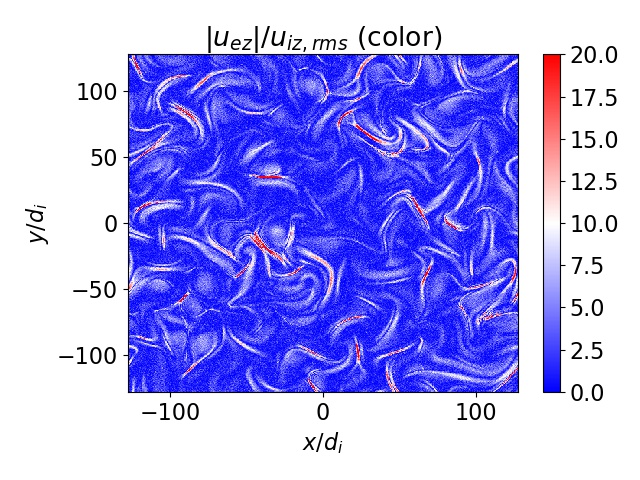}
    \put(-225,160){\large (a)}\\
    \includegraphics[clip=true,trim=0.25cm 0.7cm 0.25cm 0.7cm,width=0.49\linewidth]{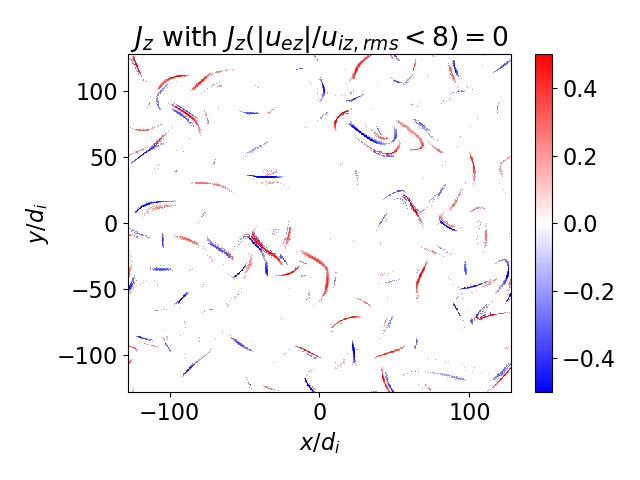}
    \put(-225,160){\large (b)}
    \caption{Ratio $|u_{ez}|/u_{iz,rms}$ of magnitude of parallel electron bulk velocity $|u_{ez}|$ and root-mean-square value of parallel ion bulk velocity $u_{iz,rms}$ (a)  and current density whose values are set to zero wherever $|u_{ez}|/u_{iz,rms} <8$ (b) in the full simulation domain at $\omega_{ci}t=50$. \label{fig:uezDuiz}}
    \end{center}
\end{figure}

In our simulations, $u_{ez}$ in current sheets is at least eight times larger than the global root-mean-square value of $u_{iz}$, $|u_{ez}|/u_{iz,rms} \gtrsim 8$ (Fig. \ref{fig:uezDuiz}a). In order to check if the condition $|u_{ez}|/u_{iz,rms} \gtrsim 8$ in our simulations is specific to current sheets, we set to zero the values of $J_z$ at the grid points where   the condition is not satisfied, i.e., where $|u_{ez}|/u_{iz,rms} < 8$. The non-zero values of the so conditioned $J_z$, plotted in Fig. \ref{fig:uezDuiz}b, correspond to the locations where the condition is satisfied and fall primarily in current sheets endorsing the specificity of the condition to current sheets formed in our simulations. Note that the number on the RHS of the inequality $|u_{ez}|/u_{iz,rms} \gtrsim 8$ is not universal for collisionless plasma turbulence but is specific to the parameters of our simulations. This number, however, would always be much greater than unity as long as the current sheets thin down to below ion inertial length.

Perpendicular electron bulk velocity, shown in Fig. \ref{fig:ueperp_uiperp_t50}a, also develops gradients in and around current sheets.  
In a sharp contrast to the parallel component, its magnitude is  almost equal to the magnitude of the perpendicular ion bulk velocity. The difference in the two, shown in Fig. \ref{fig:ueperp_uiperp_t50}b, is noticeable only around current sheets. Line-outs in Fig. \ref{fig:xlineouts}b show that significant difference in the  magnitudes of the two occur  where
$|\mathbf{u}_{e\perp}|$ has relatively sharper variation.

A measure of perpendicular  shear flows is parallel flow vorticity which develop near current sheets for both the ion and electron flows, as shown Fig. \ref{fig:ueperp_uiperp_t50}c and \ref{fig:ueperp_uiperp_t50}d. Development of parallel ion vorticity near current sheets has been observed in other particle-in-cell (PIC) and PIC-hybrid simulations \citep{franci2015,parashar2016}. Our simulations, on the other hand, show development of parallel electron vorticity near current sheets. The two vorticities are of the  same order of magnitude with electron vorticity typically larger than the ion vorticity (Fig. \ref{fig:xlineouts}c).

\begin{figure*}
  \begin{center}
    \includegraphics[clip=true,trim=0.3cm 1.0cm 0.8cm 0.8cm,width=0.45\linewidth]{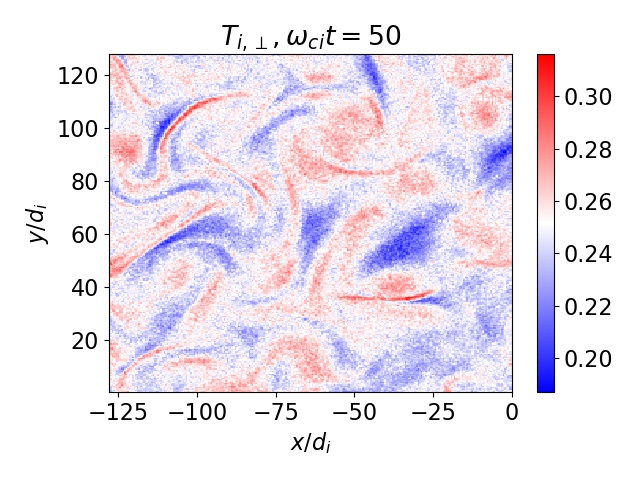}
    \put(-220,150){\large (a)}
    \hspace{0.2in}
  \includegraphics[clip=true,trim=0.3cm 1.0cm 0.8cm 0.8cm,width=0.45\linewidth]{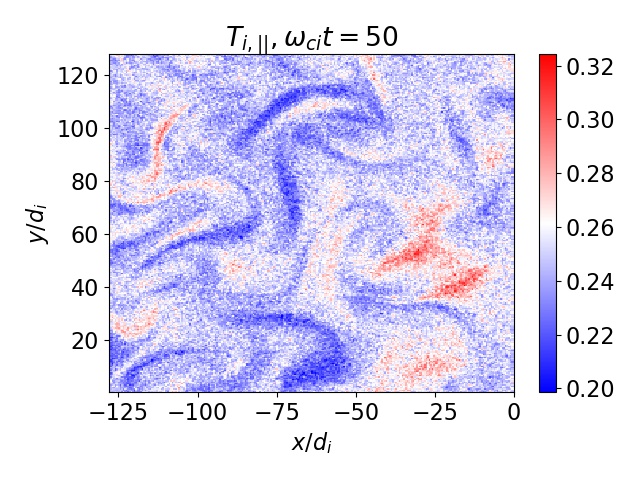}
  \put(-220,150){\large (b)}
  \vspace{0.1in}
    \includegraphics[clip=true,trim=0.3cm 1.0cm 0.8cm 0.8cm,width=0.45\linewidth]{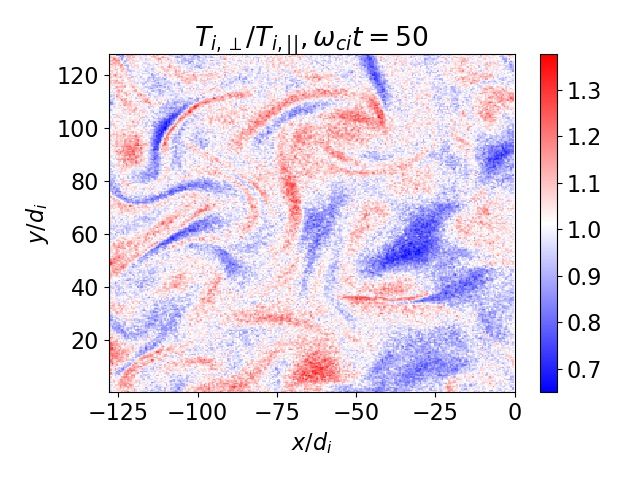}
    \put(-220,150){\large (c)}
      \hspace{0.2in}
      \includegraphics[clip=true,trim=0.3cm 1.0cm 0.8cm 0.8cm,width=0.45\linewidth]{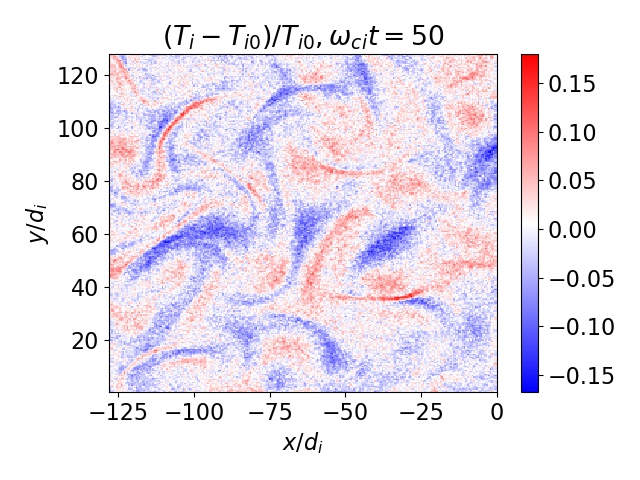}
  \put(-220,150){\large (d)}
  \caption{Perpendicular (a) and parallel (b) ion temperatures ($T_{i,\perp}$ and $T_{i,||}$), ion temperature anisotropy $T_{i,\perp}/T_{i,||}$ (c) and fractional change $(T_i-T_{i0})/T_{i0}$ in total ion temperature $T_i=(2T_{i,\perp}+T_{i,||})/3$ with respect to its initial value $T_{i0}$ (d) shown in  the top-left quadrant of the full simulation domain at $\omega_{ci}t=50$.  \label{fig:temp_aniso_t50}}
    \end{center}
\end{figure*}

\begin{figure*}
  \begin{center}
    \hspace{0.01in}
    \includegraphics[clip=true,trim=0.45cm 1.45cm 0.75cm 0.5cm,width=0.39\linewidth]{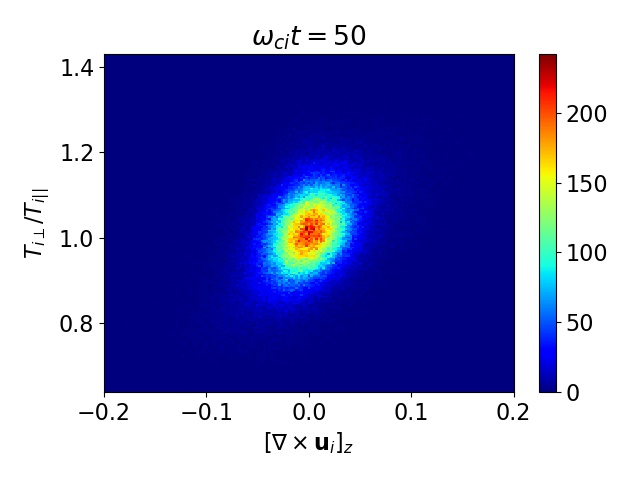}
  \put(-155,100){\large {\color{white}(a)}}
  \hspace{0.57in}
  \includegraphics[clip=true,trim=0.45cm 1.45cm 0.75cm 0.5cm,width=0.39\linewidth]{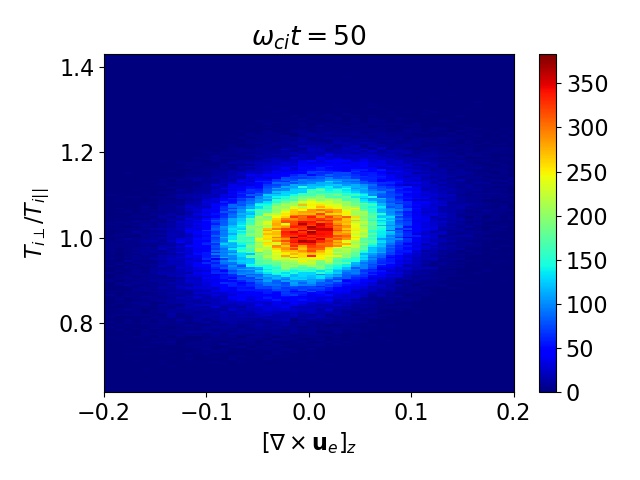}
    \put(-155,100){\large {\color{white}(b)}}\\
  \includegraphics[clip=true,trim=0.45cm 1.45cm 0.75cm 1.5cm,width=0.4\linewidth]{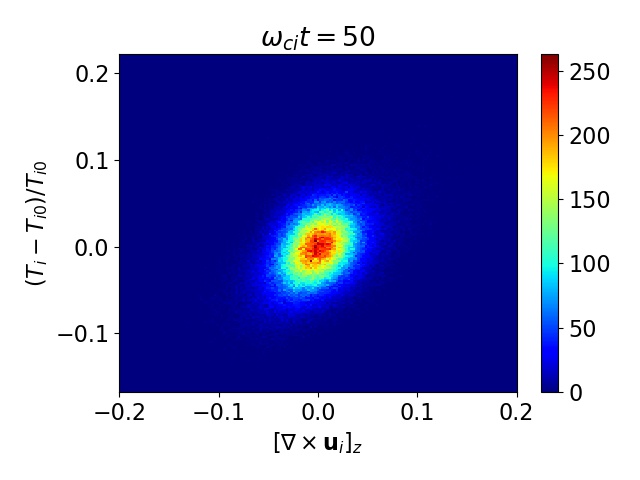}
  \put(-155,100){\large {\color{white}(c)}}
  \hspace{0.5in}
  \includegraphics[clip=true,trim=0.45cm 1.45cm 0.75cm 1.5cm,width=0.4\linewidth]{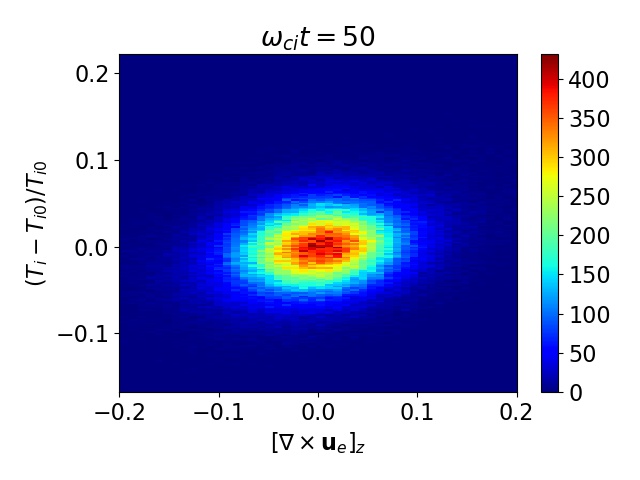}
  \put(-155,100){\large {\color{white}(d)}}\\
  \includegraphics[clip=true,trim=0.45cm 1.45cm 0.75cm 1.5cm,width=0.4\linewidth]{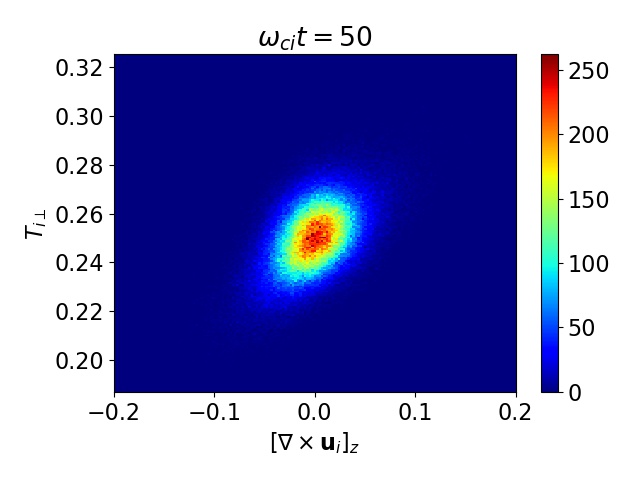}
  \put(-150,100){\large {\color{white}(e)}}
  \hspace{0.5in}
  \includegraphics[clip=true,trim=0.45cm 1.45cm 0.75cm 1.5cm,width=0.4\linewidth]{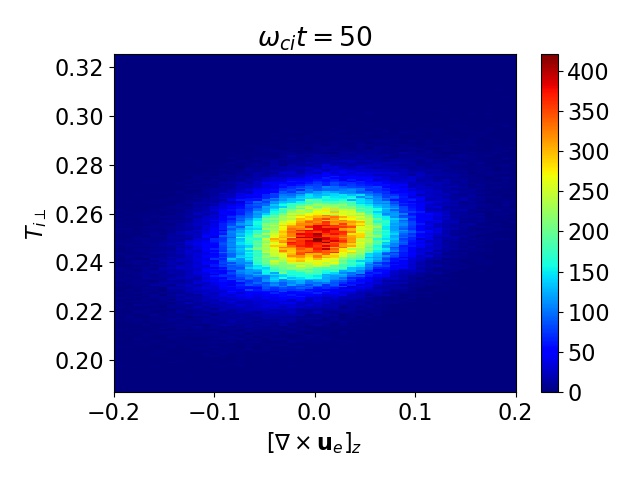}
  \put(-150,100){\large {\color{white}(f)}}\\
  \includegraphics[clip=true,trim=0.45cm 0.45cm 0.75cm 1.5cm,width=0.4\linewidth]{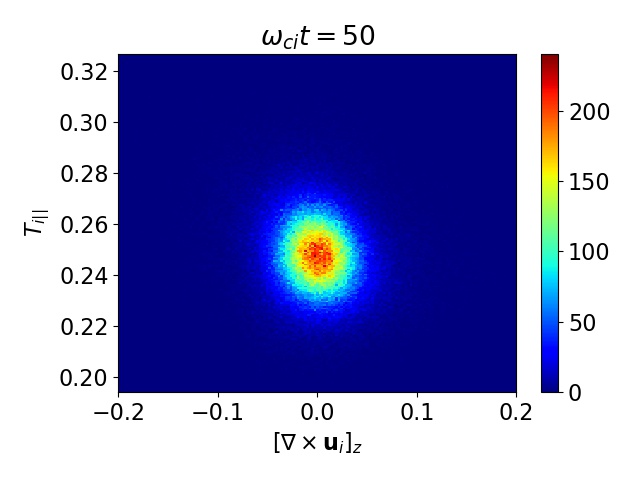}
  \put(-150,100){\large {\color{white}(g)}}
  \hspace{0.5in}
  \includegraphics[clip=true,trim=0.45cm 0.45cm 0.75cm 1.5cm,width=0.4\linewidth]{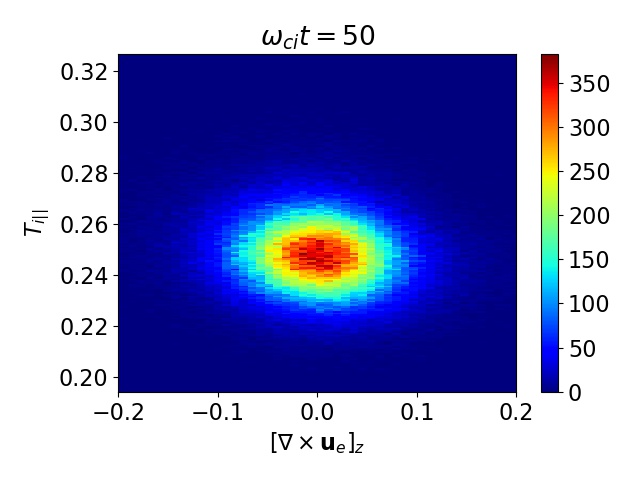}
  \put(-150,100){\large {\color{white}(h)}}
  \caption{Two dimensional histograms of ion temperature anisotropy $T_{i,\perp}/T_{i,||}$ (first row), fractional change in ion temperature $(T_i-T_{i0})/T_{i0}$ (second row), and perpendicular (third row) and parallel (fourth row) ion temperatures ($T_{i,\perp}$ and $T_{i,||}$) with parallel ion (first column) and electron (second column) vorticities at $\omega_{ci}t=50$. \label{fig:scatter_t50}. Color represents number of counts.}
\end{center}
  \end{figure*}

By the time current sheets form, initially isotropic distribution of ions develops different temperatures parallel and perpendicular to the mean magnetic field (Fig. \ref{fig:temp_aniso_t50}a and \ref{fig:temp_aniso_t50}b). This results in development of both the parallel ($T_{i,\perp}/T_{i,||} < 1$) and perpendicular ($T_{i,\perp}/T_{i,||} > 1$) ion temperature anisotropy in the turbulence (Fig. \ref{fig:temp_aniso_t50}c). Ions have also undergone both heating and cooling by $\omega_{ci}t=50$ (Fig. \ref{fig:temp_aniso_t50}d). The regions of heating/cooling and perpendicular/parallel anisotropy  are structured mostly around current sheets consistent with other PIC and PIC-hybrid simulations \citep{franci2016,wan2015}. 
   Vlasov-hybrid simulations, free from particle noise inherent in PIC method and thus allowing a better accuracy in the calculation of the velocity moments of the ion distribution function, also show development of not only temperature anisotropy but also other non-Maxwellian features including nonzero skewness (heat flux) and high/low kurtosis concentrated in sheet-like magnetic structures with scale size of the order of an ion inertial length \citep{greco2012,servidio2012}. In these simulations of collisionless plasma turbulence  carried out in 2D-3V geometry (two dimensions in physical space and three in velocity space) for $\beta_i=2$, ion temperature anisotropy gets up to the value $\sim 1.3$, similar to what is observed in our PIC-hybrid simulations with $\beta_i=0.5$ (see Fig. \ref{fig:temp_aniso_t50}c).

 Fig. \ref{fig:scatter_t50}a, \ref{fig:scatter_t50}c and \ref{fig:scatter_t50}e respectively show that ion temperature anisotropy, change in ion temperature from its initial value, and perpendicular ion temperature are  positively correlated with parallel ion flow vorticity. Parallel ion temperature, on the other hand, seems to have relatively weak negative correlation with the parallel ion flow vorticity (Fig. \ref{fig:scatter_t50}g). These correlations with the parallel ion flow vorticity have earlier been reported in PIC- and Vlasov-hybrid simulations \citep{servidio2012,greco2012,franci2016}.
We additionally found here that the three quantities, viz., ion temperature anisotropy, change in ion temperature from its initial value, and perpendicular ion temperature, are positively  correlated with parallel electron flow vorticity as well (Fig. \ref{fig:scatter_t50}b, \ref{fig:scatter_t50}d and \ref{fig:scatter_t50}f). Parallel ion temperature, however, does not seem to have correlation with the electron flow vorticity (Fig. \ref{fig:scatter_t50}h).       

 The association of ion temperature anisotropy with the parallel ion vorticity is understood to be due to the $d_i$-scale spatial inhomogeneity of  ion shear flow which generates pressure anisotropies via pressure-strain interaction in the plane perpendicular to the magnetic field \citep{sarto2016}. Two and half dimensional fully kinetic PIC simulations of collisionless plasma turbulence carried out for $\beta_i=\beta_e=0.1$ suggest the role of traceless pressure-strain interaction, which is strongest around current sheets formed in the turbulence, in anisotropic ion heating around current sheets  \citep{yang2017a,yang2017b}. Observations by Magnetospheric Multiscale Mission in Earth's magnetosheath also confirmed that the pressure-stress interactions can convert flow energy into internal energy \citep{chasapis2018}. The association of ion temperature anisotropy with the parallel electron vorticity, on the other hand, could be due to the reason that parallel vorticities of electrons and ions are concentrated in almost the same spatial regions  (see Fig. \ref{fig:ueperp_uiperp_t50}c and \ref{fig:ueperp_uiperp_t50}d). The co-location of electron and ion vorticities was also observed in PIC simulations of collisionless plasma turbulence \citep{yang2017a,yang2017b}. These PIC simulations, however, also show co-location of the electron vorticity and electron pressure-strain interaction physics of which is absent in our simulations.

Note that earlier simulations reported the development of ion temperature anisotropy and its correlation with ion flow vorticity at the time of the maximum turbulent activity, which was taken to be the time when RMS value of parallel current density peaks \citep{yang2017a,yang2017b,servidio2012,greco2012,franci2016}.    Our results presented here, however, show that ion temperature anisotropy exhibits similar behavior at the time when current sheets have just formed ($\omega_{ci}t=50$), much earlier than the time of maximum turbulent activity    ($\omega_{ci}t=150$ in our simulations, see Fig. \ref{fig:rms_evolution}).
 This observation leads to the speculation that  the processes of current sheet formation arising from turbulent cascade might themselves contribute to ion heating/cooling and development of ion temperature anisotropy in collisionless plasma turbulence, since such a correlation does not exist in the initial conditions. The formation of current sheet is associated with the generation of quadrupole vorticity structure \citep{parashar2016} in which pressure-strain interaction can produce temperature anisotropies \citep{sarto2016}. Indeed, 2.5D PIC simulations of collisionless plasma turbulence show that ion pressure-strain interaction term is finite and larger than the pressure dilation term at times (during which current sheets might be forming) much before mean square current reaches its maximum (see Fig. 2 in \citep{yang2017a}).   Later, development of plasma instabilities in the formed current sheets might also contribute to anisotropic heating/cooling   \citep{daughton2004,haynes2014,karimabadi2013}.  

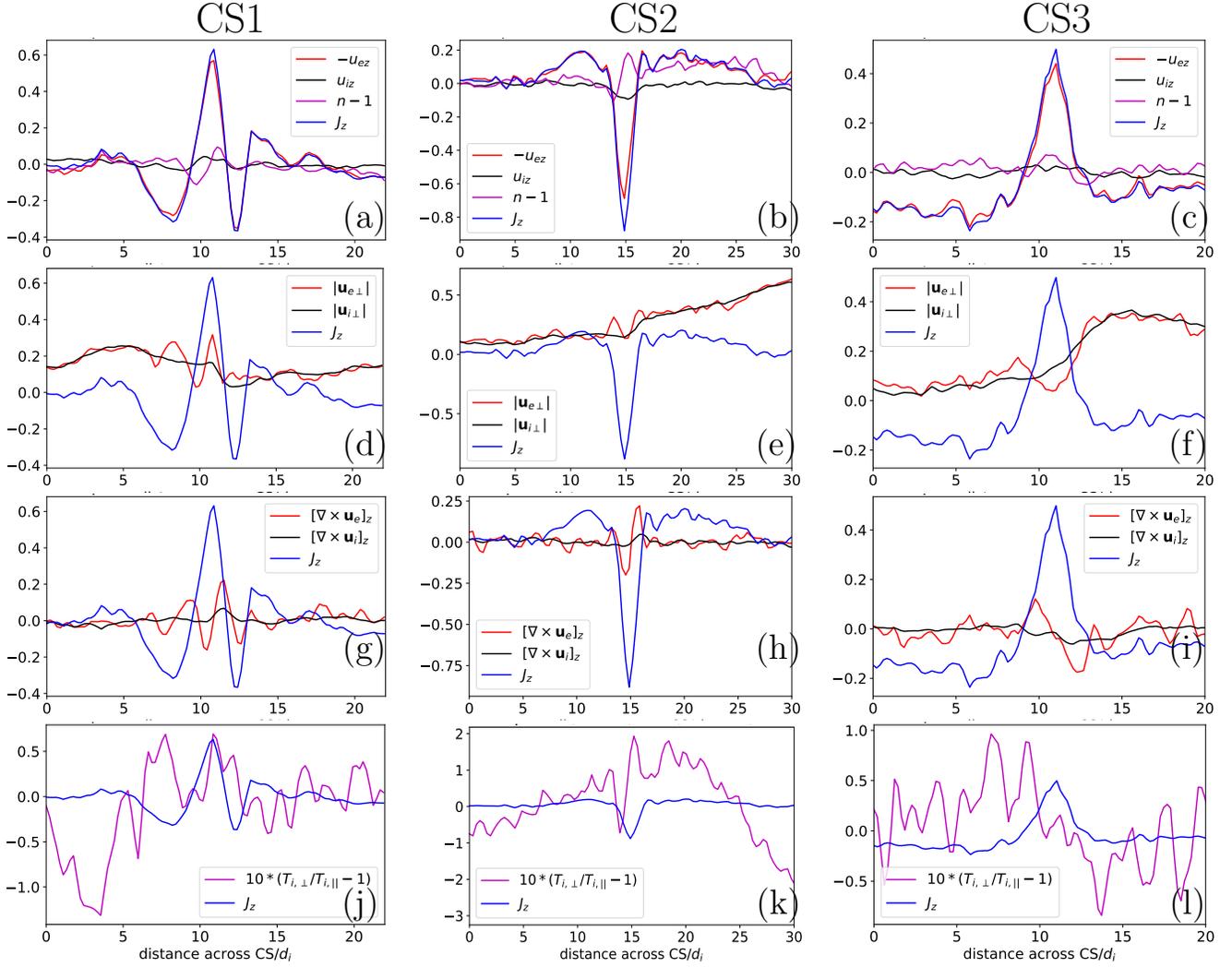
\begin{figure*}
  \begin{center}
  \includegraphics[clip=true,trim=0.0cm 1.30cm 0cm 2.0cm,width=0.33\linewidth]{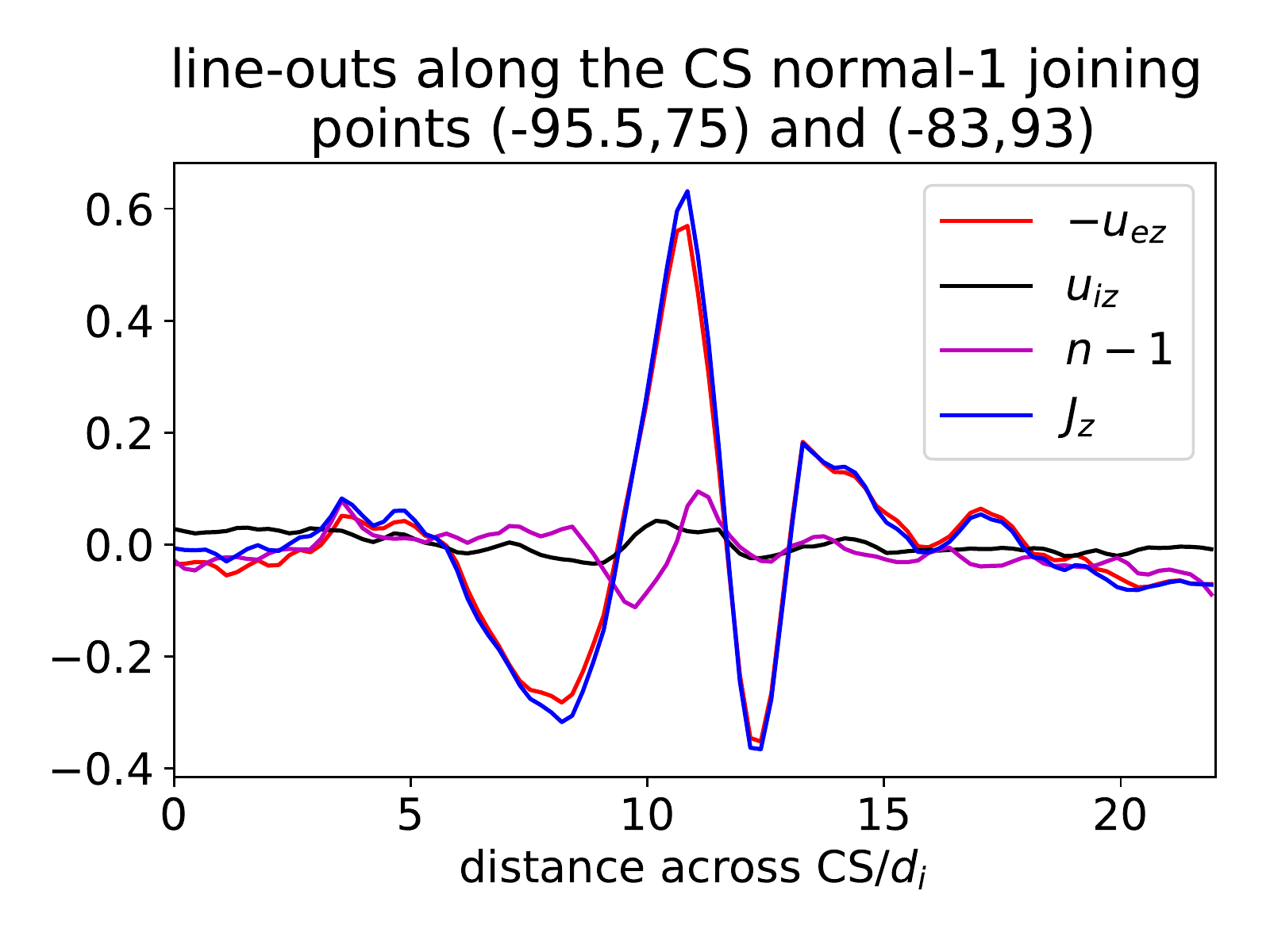}
  \put(-25,15){\large (a)}
    \put(-85,95){\Large CS1}
  \includegraphics[clip=true,trim=0.0cm 1.30cm 0cm 2.0cm,width=0.33\linewidth]{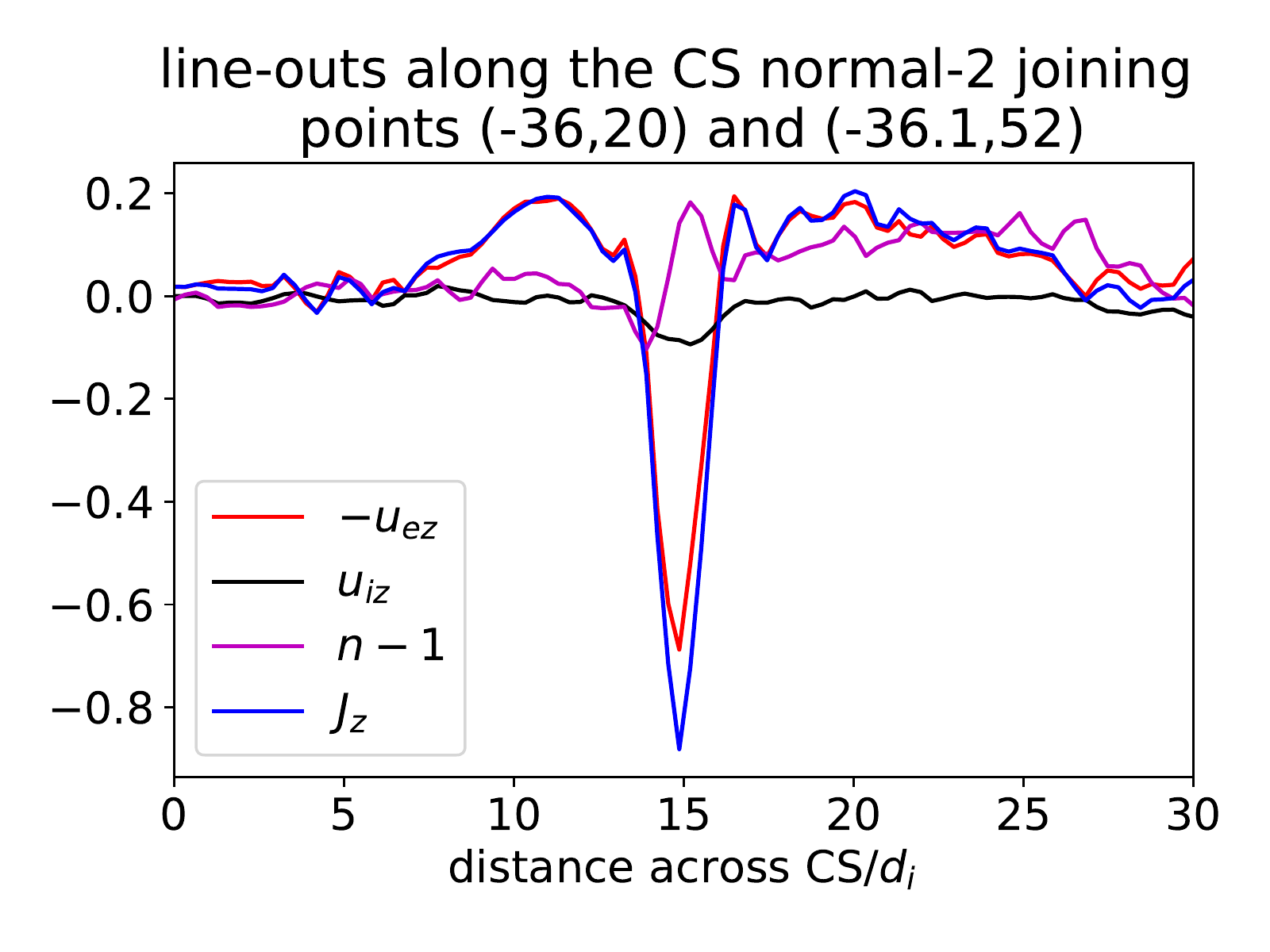}
  \put(-25,15){\large (b)}
      \put(-85,95){\Large CS2}
  \includegraphics[clip=true,trim=0.0cm 1.30cm 0cm 2.0cm,width=0.33\linewidth]{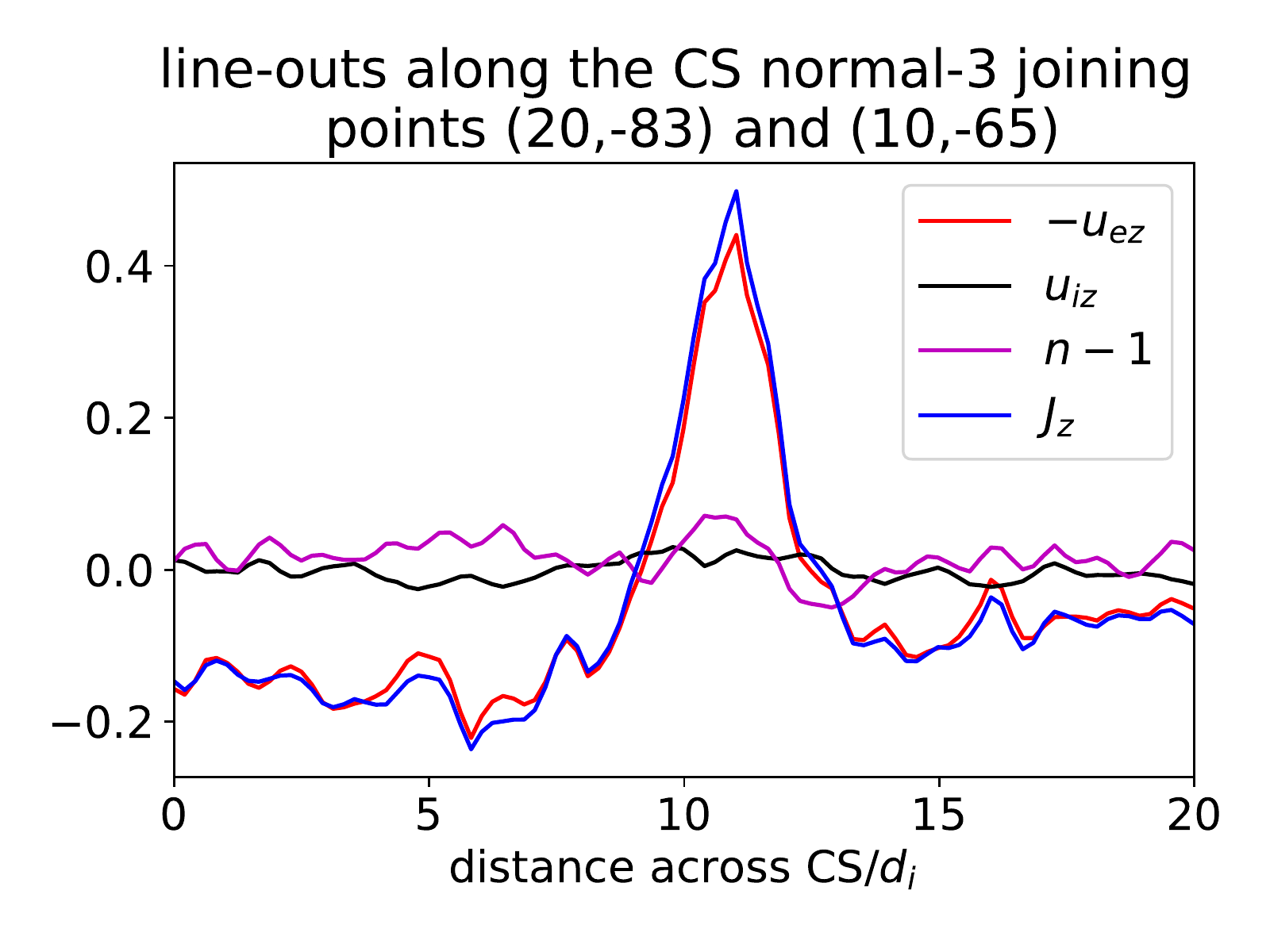}
  \put(-25,15){\large (c)}
      \put(-85,95){\Large CS3}\\
      \includegraphics[clip=true,trim=0.0cm 1.30cm 0cm 2.0cm,width=0.33\linewidth]{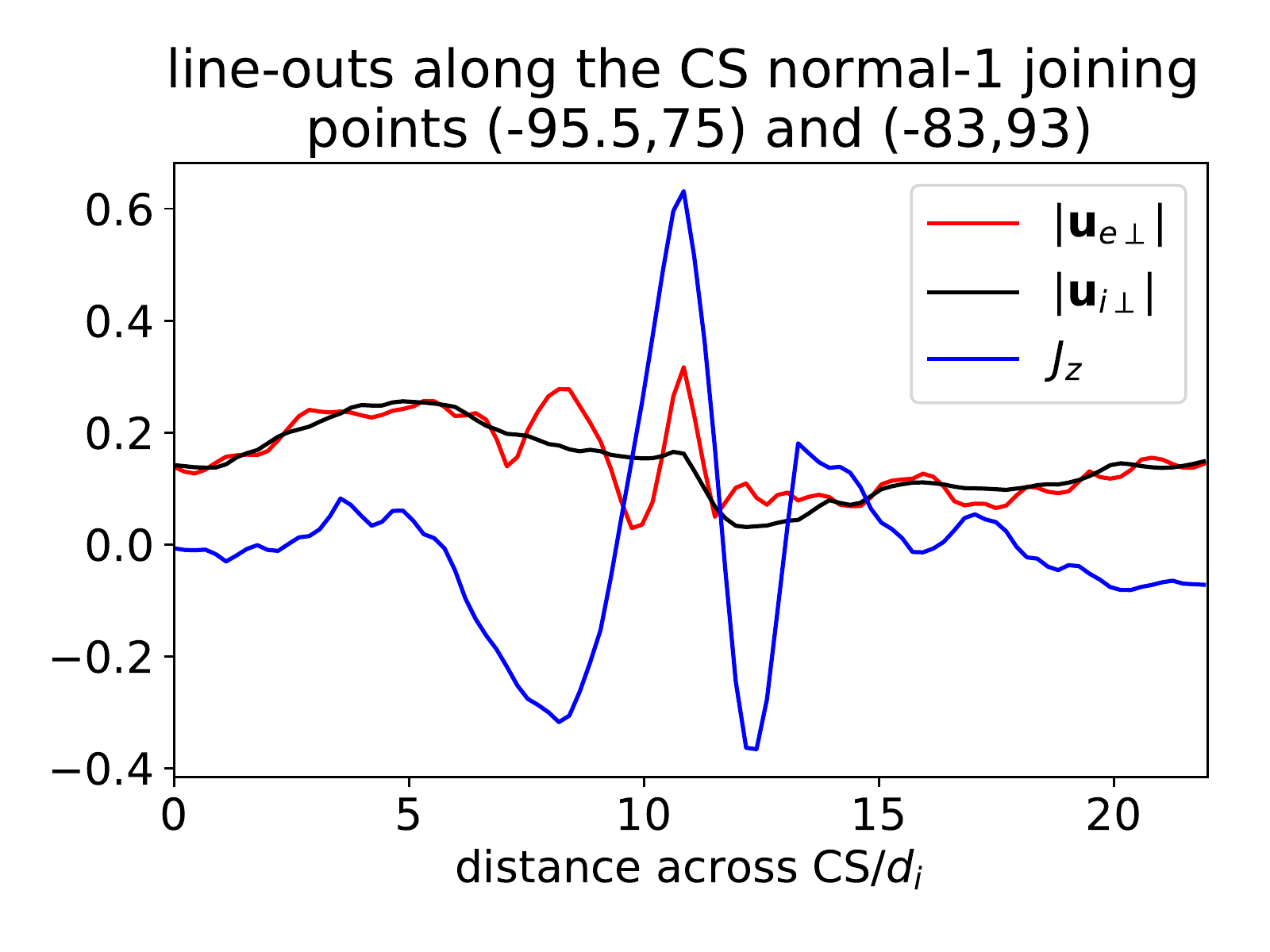}
  \put(-25,15){\large (d)}
  \includegraphics[clip=true,trim=0.0cm 1.30cm 0cm 2.0cm,width=0.33\linewidth]{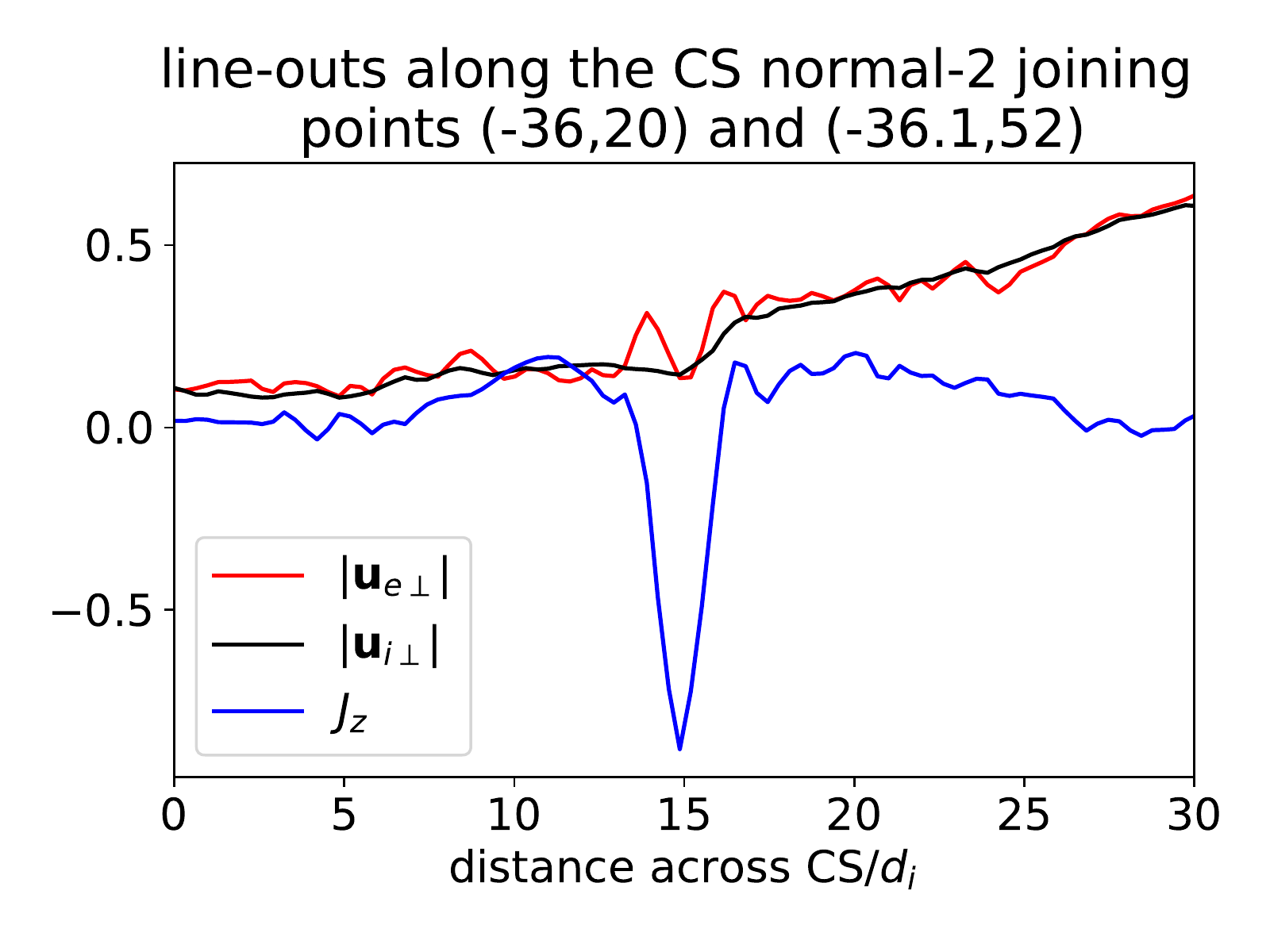}
  \put(-25,15){\large (e)}
  \includegraphics[clip=true,trim=0.0cm 1.30cm 0cm 2.0cm,width=0.33\linewidth]{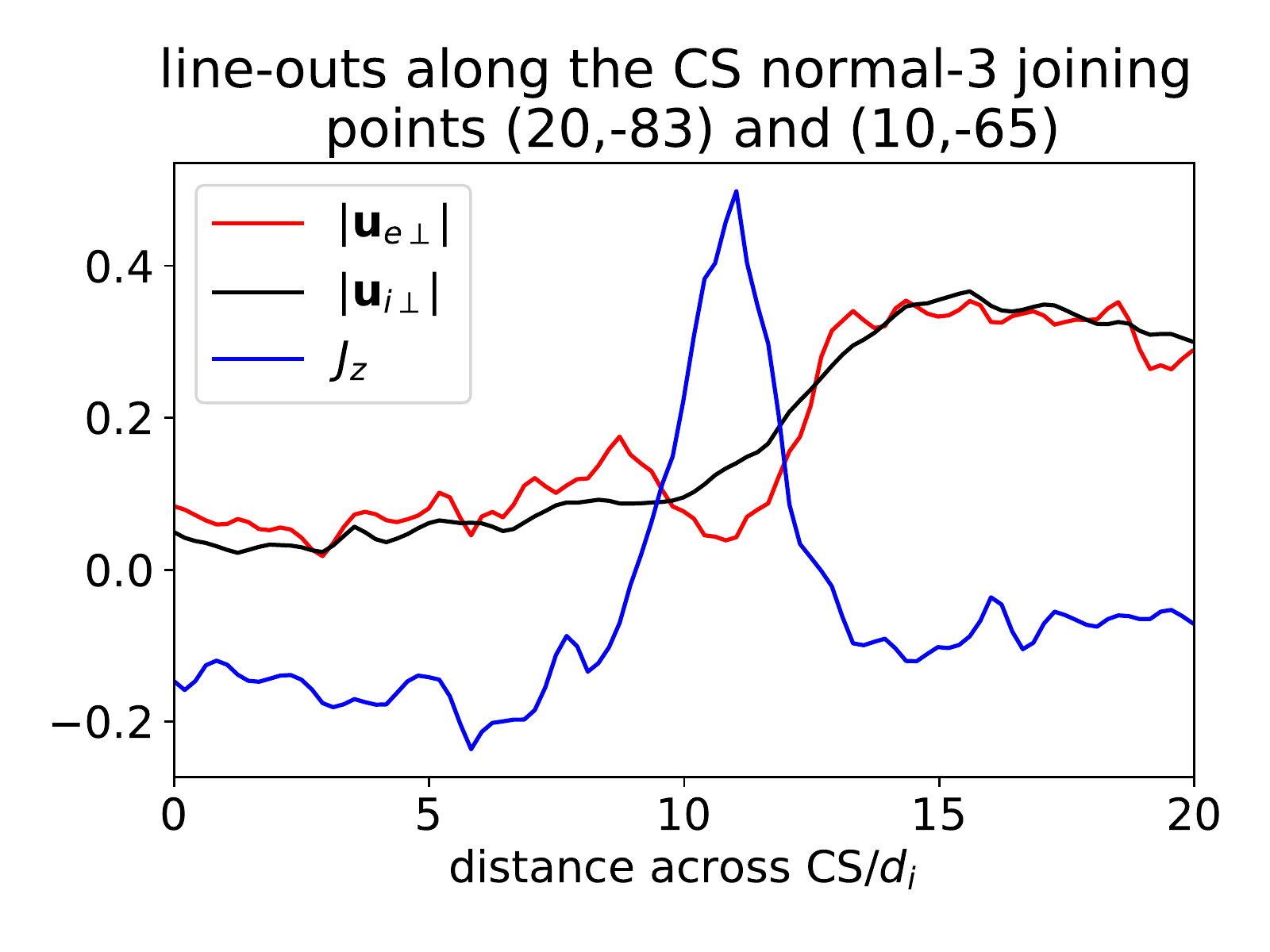}
  \put(-25,15){\large (f)}\\
      \includegraphics[clip=true,trim=0.0cm 1.30cm 0cm 2.0cm,width=0.33\linewidth]{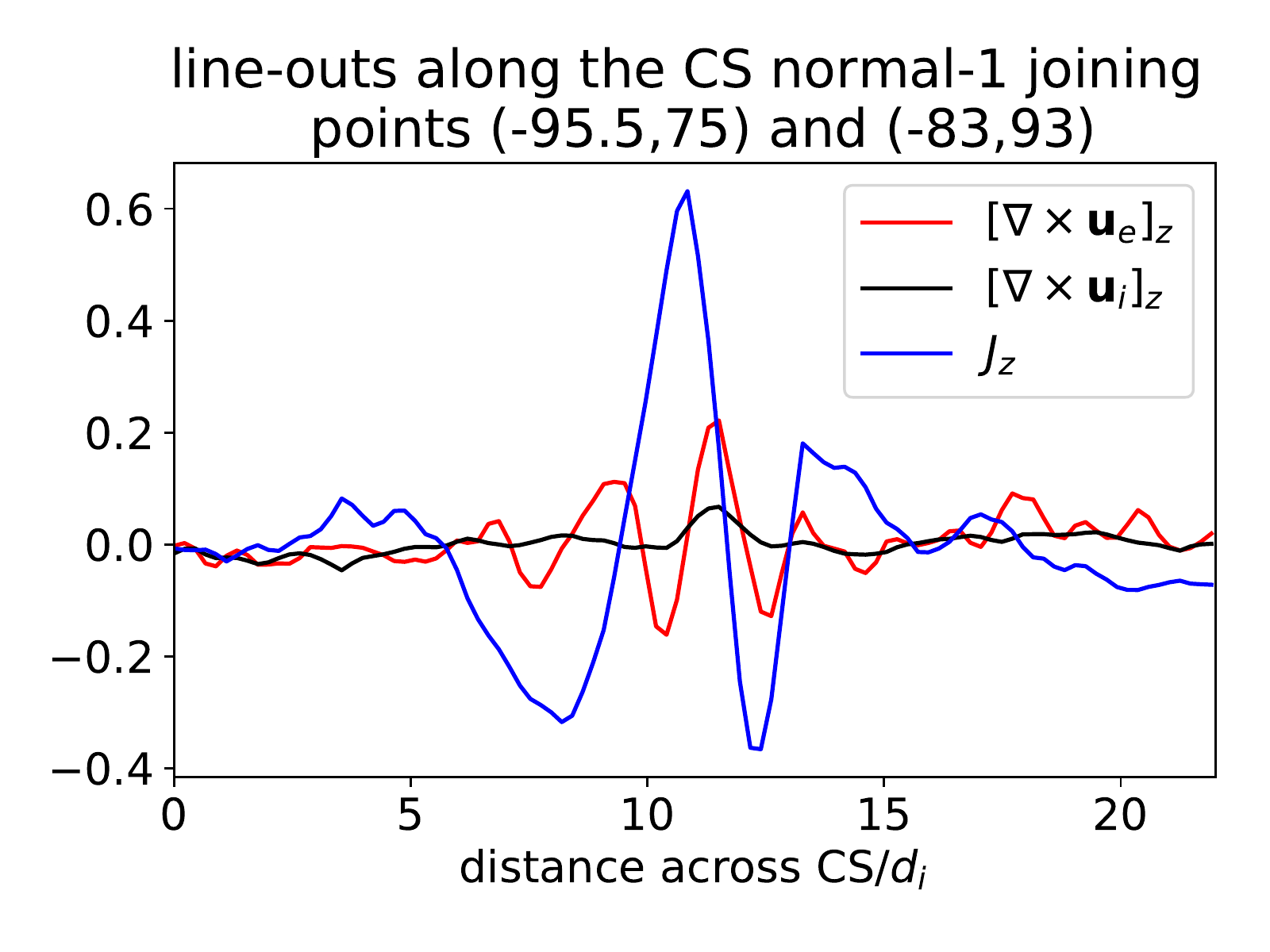}
  \put(-25,25){\large (g)}
      \includegraphics[clip=true,trim=0.0cm 1.30cm 0cm 2.0cm,width=0.33\linewidth]{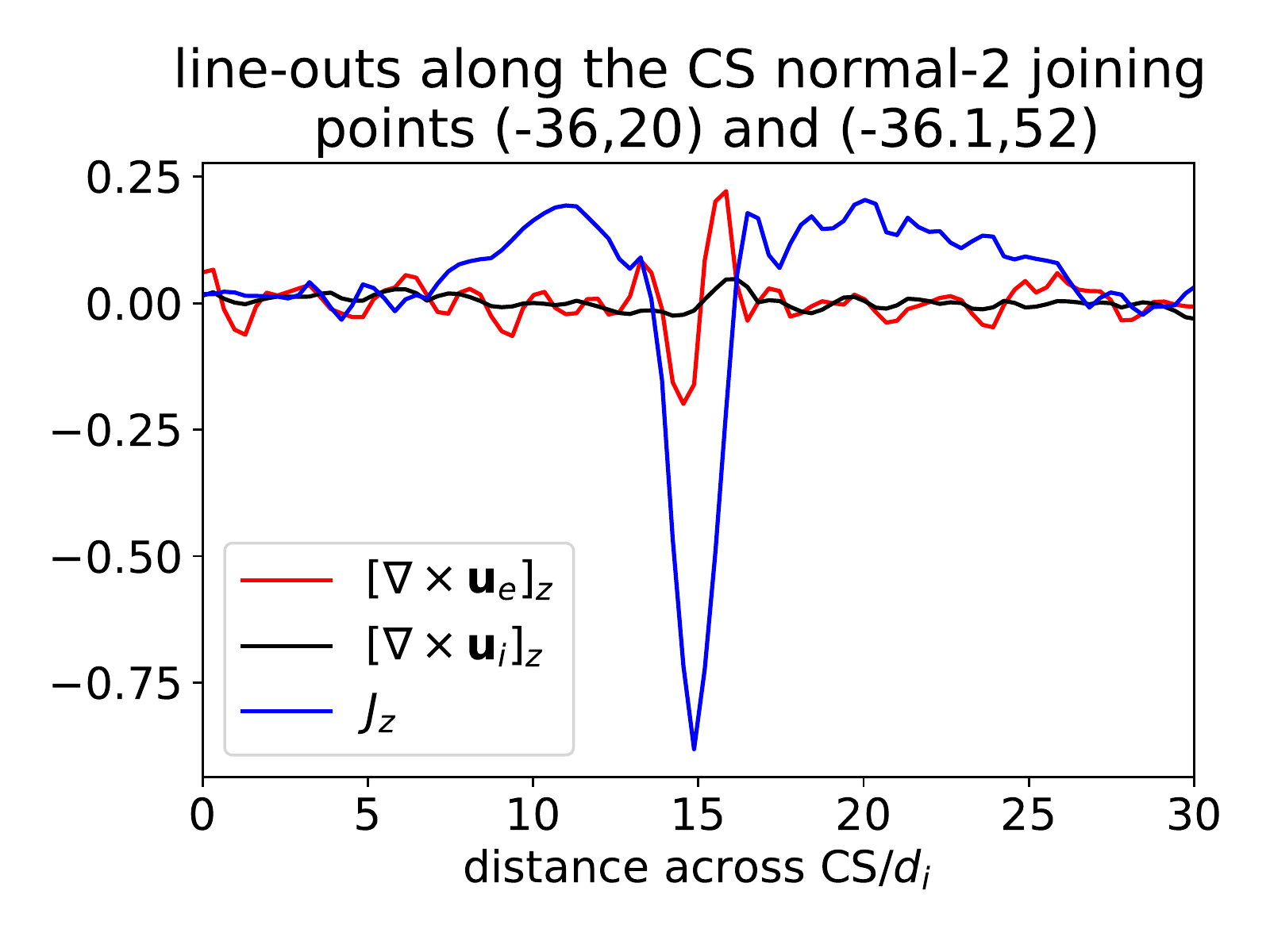}
  \put(-25,25){\large (h)}
      \includegraphics[clip=true,trim=0.0cm 1.30cm 0cm 2.0cm,width=0.33\linewidth]{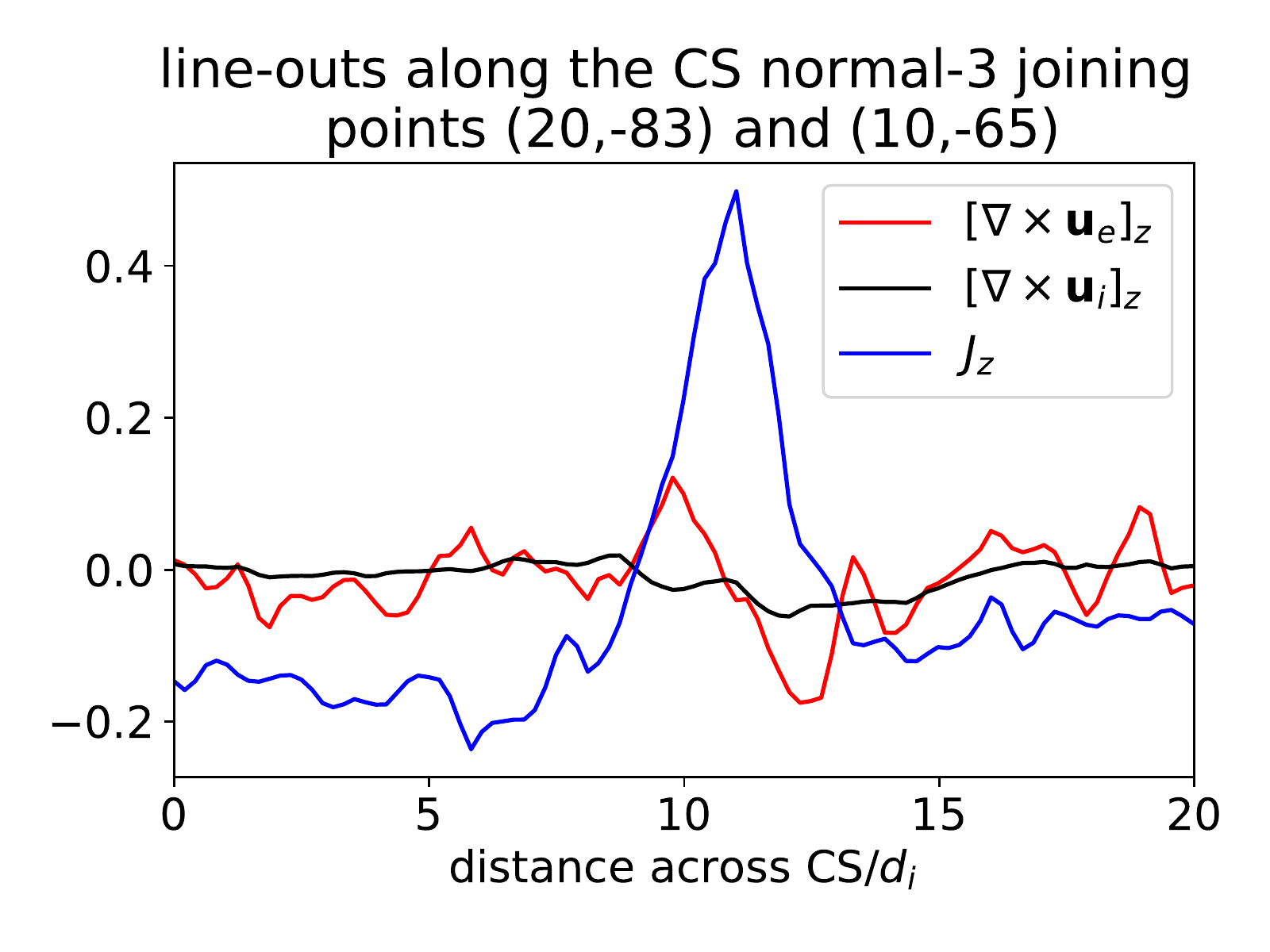}
      \put(-25,25){\large (i)}\\
      \includegraphics[clip=true,trim=0.0cm 0.25cm 0cm 2.0cm,width=0.33\linewidth]{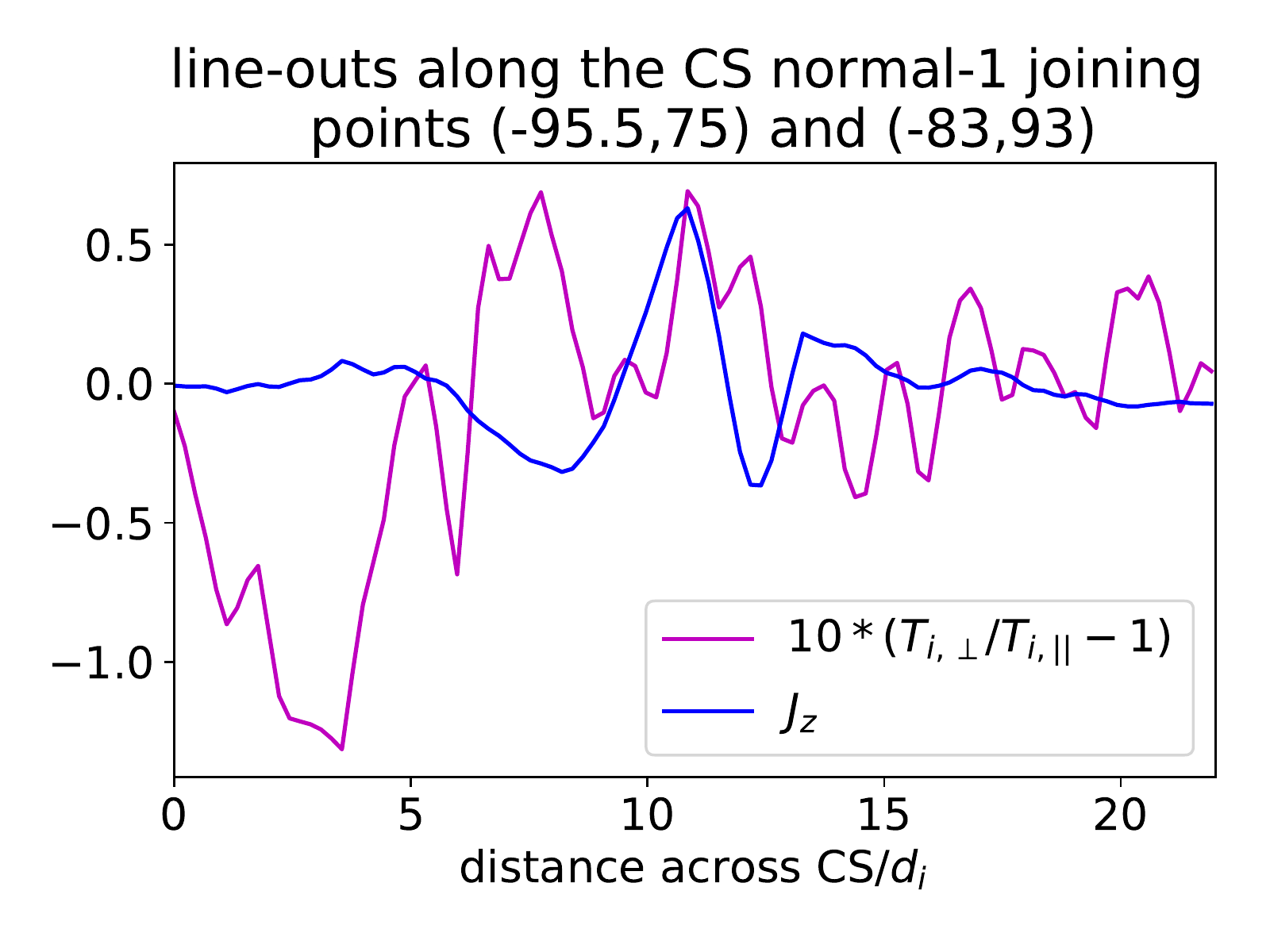}
  \put(-25,25){\large (j)}
     \hspace{0.1in} \includegraphics[clip=true,trim=0.0cm 0.25cm 0cm 2.0cm,width=0.31\linewidth,height=0.20\linewidth]{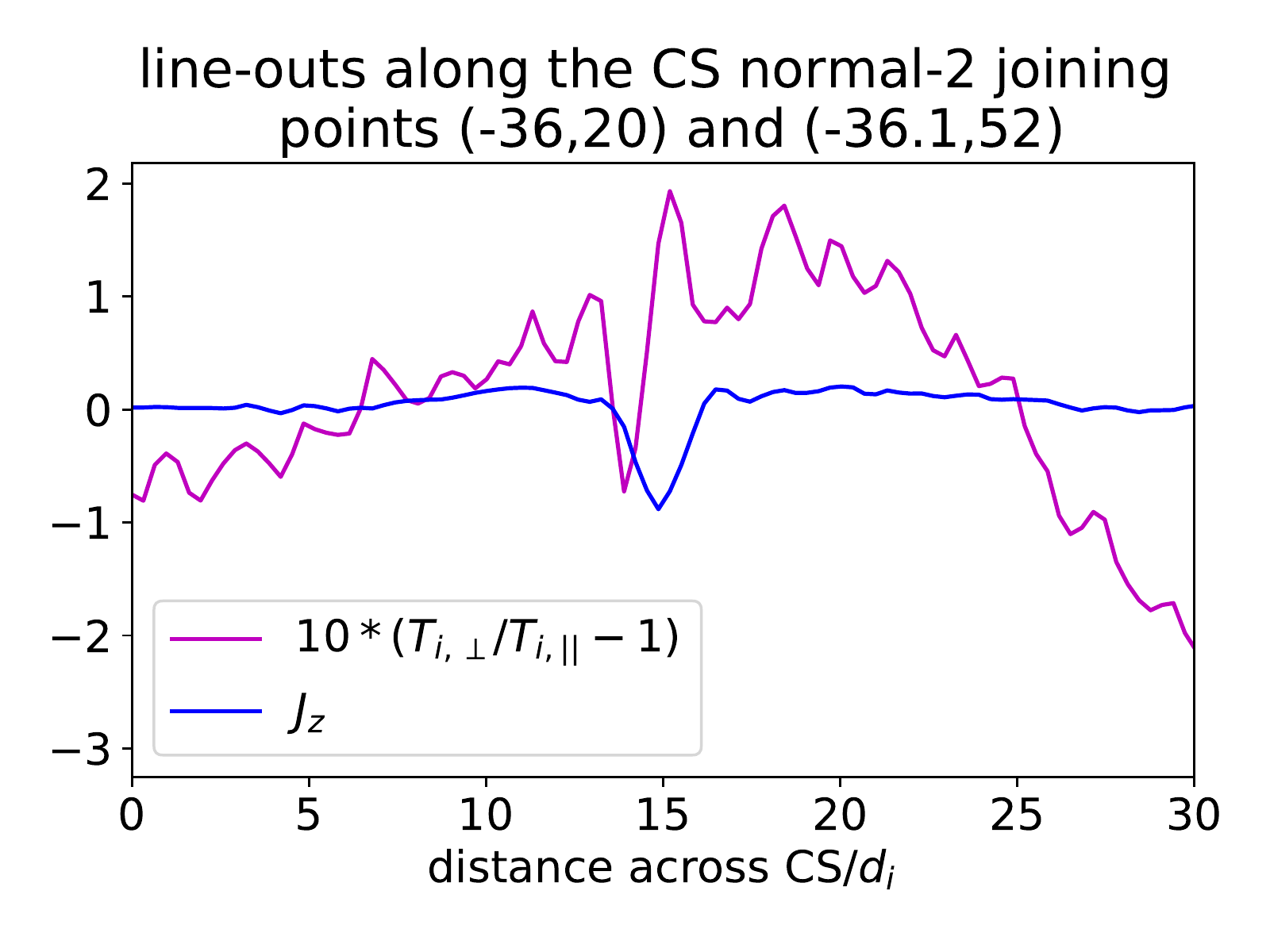}
  \put(-25,25){\large (k)}
      \includegraphics[clip=true,trim=0.0cm 0.25cm 0cm 2.0cm,width=0.33\linewidth]{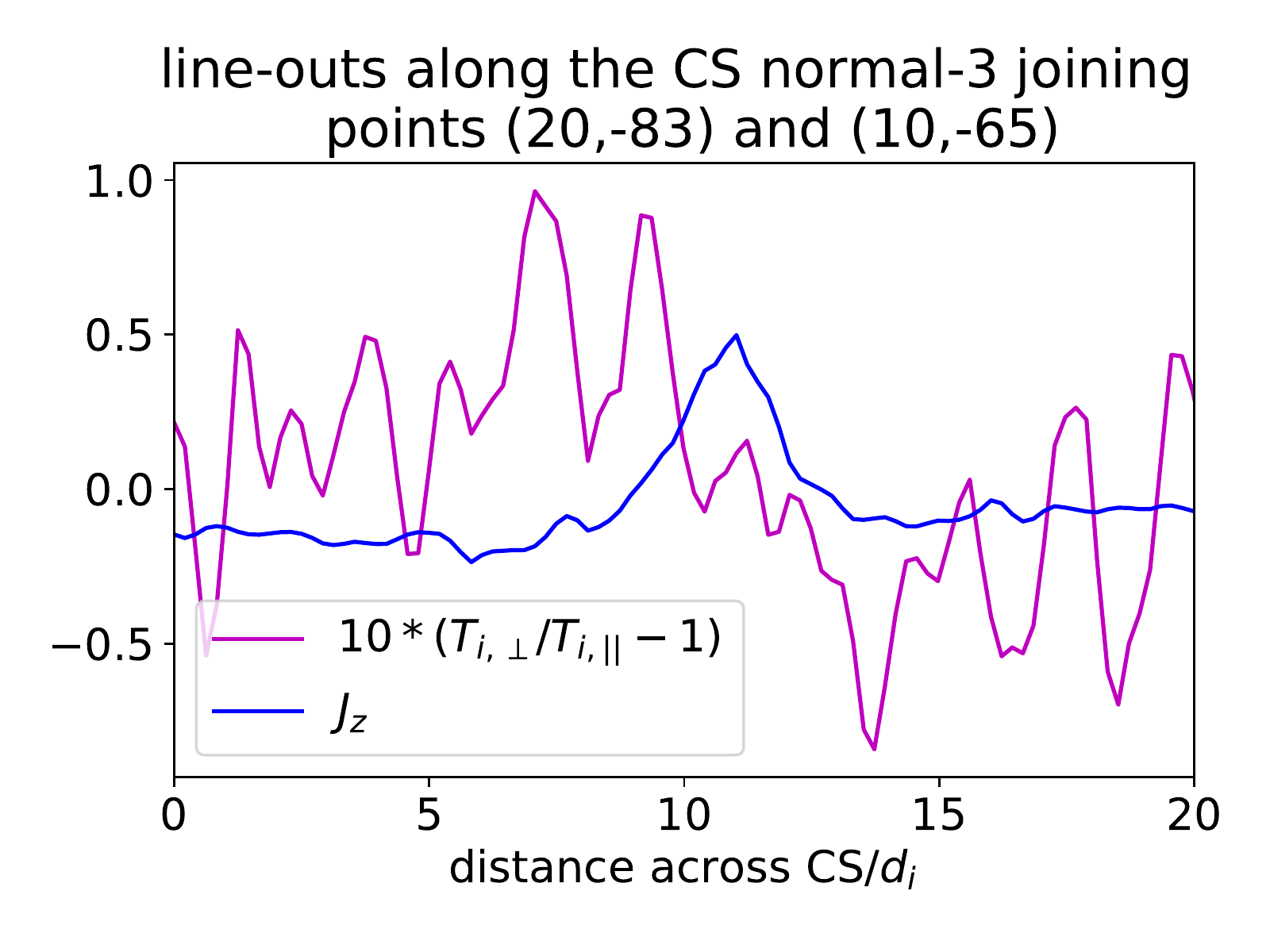}
      \put(-25,25){\large (l)}
      \caption{Line-outs along the current sheet normals numbered 1 (left column, CS1), 2 (middle column, CS2) and 3 (right column, CS3) in Fig. \ref{fig:jz_t50_t150} at $\omega_{ci}t=50$. Line-outs of  -$u_{ez}/v_{Ai}$, $u_{iz}/v_{Ai}$, $n/n_0-1$ (first row), $|\mathbf{u}_{e\perp}|/v_{Ai}$,  $|\mathbf{u}_{i\perp}|/v_{Ai}$ (second row), $\omega_{ci}^{-1}[\nabla \times \mathbf{u}_e]_z$, $\omega_{ci}^{-1}[\nabla \times \mathbf{u}_i]_z$ (third row) and deviation from temperature isotropy $10 \times (T_{i,\perp}/T_{i,||}-1)$ (fourth row). Electron and  quantities are plotted by red and black lines, respectively. Line-out of $J_z/n_0ev_{Ai}$ is plotted by blue line in all the sub-plots. \label{fig:lineouts_across_CS_t50}}
      \end{center}
\end{figure*}

We now turn our attention to individual current sheets. We select three current sheets, numbered 1-3 in Fig. \ref{fig:jz_t50_t150}, from the full simulation domain based on the criteria that they are relatively isolated from neighboring current sheets so that the features of an individual current sheet are discernible. 
Line-outs of various electron and ion quantities along the current sheet normals are shown in   Fig. \ref{fig:lineouts_across_CS_t50}  at $\omega_{ci}t=50$. The current sheet characteristics already observed in Figs. \ref{fig:ue_ui_n_t50} and \ref{fig:ueperp_uiperp_t50} can now be appreciated in individual current sheets: return current system with dominance of parallel electron bulk velocity and relatively slow variation of plasma number density (Fig. \ref{fig:lineouts_across_CS_t50}a-\ref{fig:lineouts_across_CS_t50}c), similar magnitudes of perpendicular electron and ion bulk velocities except in current sheets (Fig. \ref{fig:lineouts_across_CS_t50}d-\ref{fig:lineouts_across_CS_t50}f), development of parallel ion and electron vorticities (Fig. \ref{fig:lineouts_across_CS_t50}g-\ref{fig:lineouts_across_CS_t50}i).

Additional details on current sheet characteristics can also be seen in the line-outs  shown in Fig. \ref{fig:lineouts_across_CS_t50}.  Magnitude of the perpendicular electron bulk velocity  (Fig. \ref{fig:lineouts_across_CS_t50}d-\ref{fig:lineouts_across_CS_t50}f) in current sheets is significantly smaller than the parallel electron bulk velocity  (Fig. \ref{fig:lineouts_across_CS_t50}a-\ref{fig:lineouts_across_CS_t50}c). Therefore electric current in the sheets flow primarily parallel to the applied magnetic field. Perpendicular electron bulk velocity  (Fig. \ref{fig:lineouts_across_CS_t50}d-\ref{fig:lineouts_across_CS_t50}f) and parallel electron vorticity  (Fig. \ref{fig:lineouts_across_CS_t50}g-\ref{fig:lineouts_across_CS_t50}i) in current sheets vary faster than their ion counterparts.
Parallel electron vorticity is larger than parallel ion vorticity in current sheets and  changes sign close to the current sheet center to have opposite peaks near the current sheet edges  (Fig. \ref{fig:lineouts_across_CS_t50}g-\ref{fig:lineouts_across_CS_t50}i). Such a behavior is not very clear for the parallel ion vorticity.

With the change in sign of parallel electron vorticity across the current sheets CS2 and CS3,  the deviation of the perpendicular to parallel ion temperature ratio from its isotropic value of unity, $T_{i,\perp}/T_{i,||}-1$, also changes sign (Fig. \ref{fig:lineouts_across_CS_t50}k and \ref{fig:lineouts_across_CS_t50}l).  This, i.e., perpendicular/parallel anisotropy for positive/negative electron vorticity, is consistent with the correlation of ion temperature anisotropy with parallel electron vorticity shown in  the scatter plot of data points from the whole simulation box in Fig. \ref{fig:scatter_t50}b,  implying that the correlation holds in current sheets. Ion temperature anisotropy in the current sheet CS1, however, only partially obey the correlation --- perpendicular anisotropy for positive electron vorticity but no parallel anisotropy for negative electron vorticity (Fig. \ref{fig:lineouts_across_CS_t50}j).  This could be because formation of current sheets in turbulence is influenced by the dynamics in the neighborhood which  is generally  different for each current sheet. Among the selected current sheets, CS1  just happens to be an odd case in which the positive correlation shown in Fig. \ref{fig:scatter_t50}b is only partially obeyed.  The positive correlation is not obeyed well outside the current sheets either. Fig. \ref{fig:lineouts_across_CS_t50}j-\ref{fig:lineouts_across_CS_t50}l also shows large parallel and perpendicular anisotropy  comparable to those in current sheets outside the current sheets where parallel electron vorticity is small. Suggested by this we conjecture that the positive correlation between ion temperature anisotropy and parallel electron vorticity is obeyed primarily in and around current sheets.  

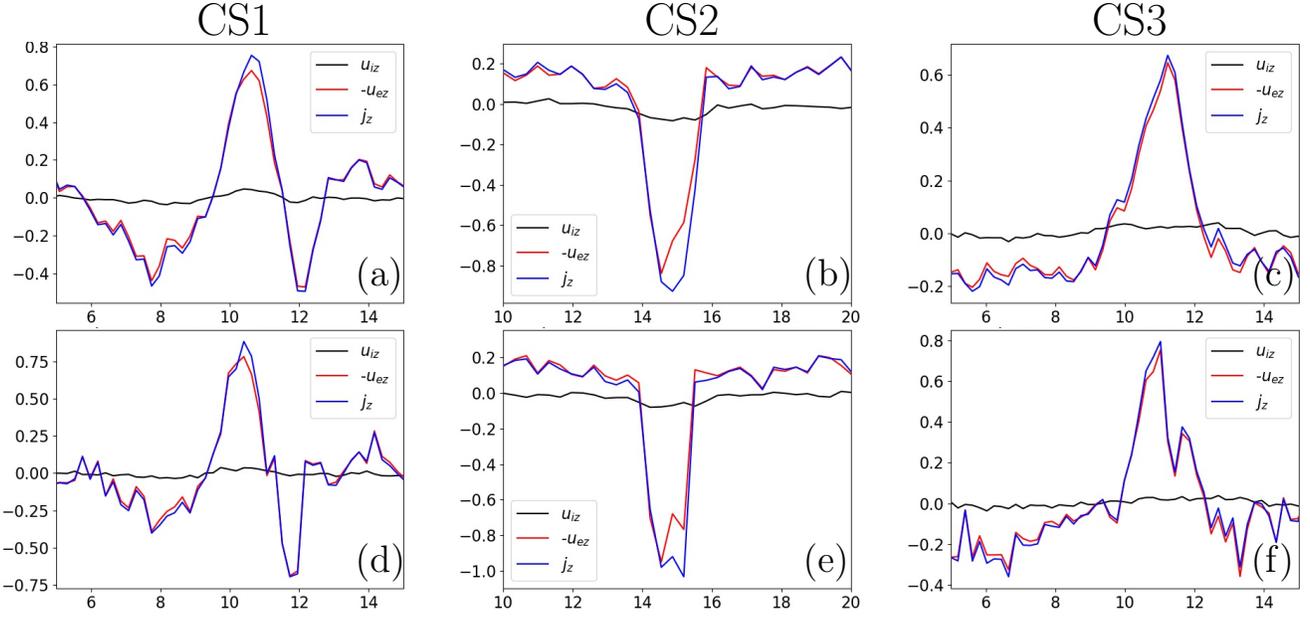
\begin{figure*}
  \begin{center}
  \includegraphics[clip=true,trim=0.0cm 0.5cm 0cm 1.4cm,width=0.33\linewidth]{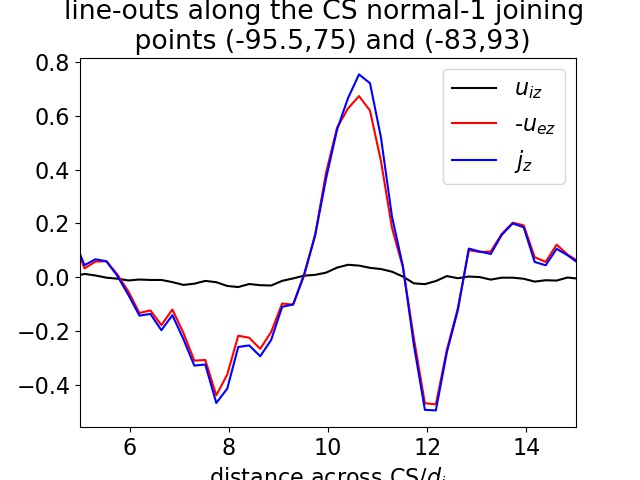}
  \put(-35,15){\large (a)}
    \put(-95,110){\Large CS1}
  \includegraphics[clip=true,trim=0.0cm 0.5cm 0cm 1.4cm,width=0.33\linewidth]{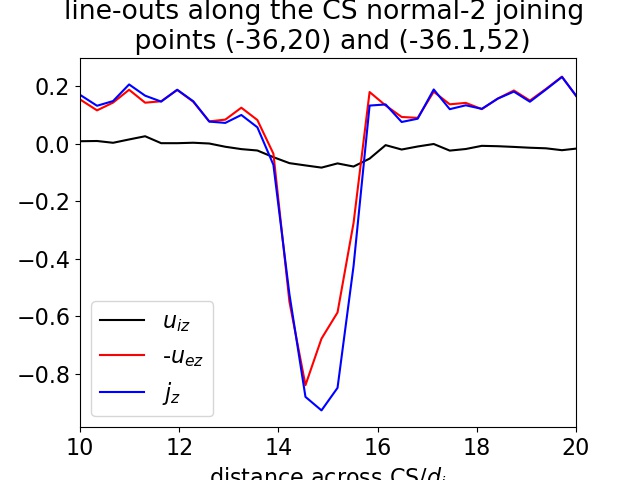}
  \put(-35,15){\large (b)}
      \put(-95,110){\Large CS2}
  \includegraphics[clip=true,trim=0.0cm 0.5cm 0cm 1.4cm,width=0.33\linewidth]{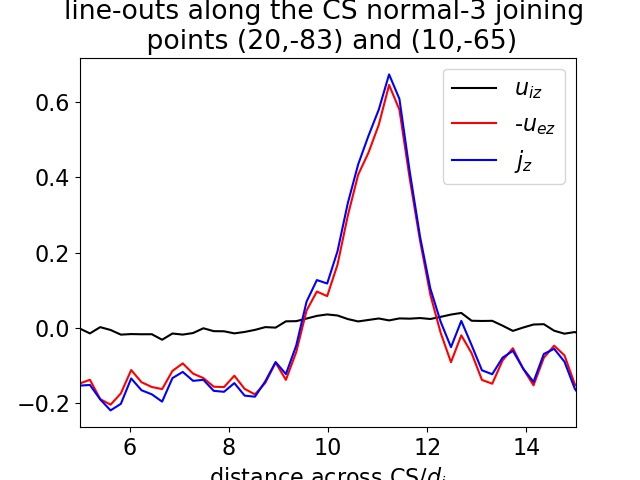}
  \put(-35,15){\large (c)}
      \put(-95,110){\Large CS3}\\
      \includegraphics[clip=true,trim=0.0cm 0.5cm 0cm 1.4cm,width=0.33\linewidth]{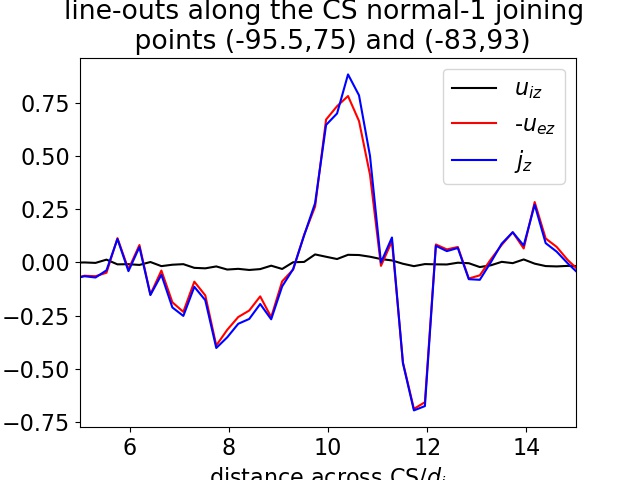}
  \put(-35,15){\large (d)}
  \includegraphics[clip=true,trim=0.0cm 0.5cm 0cm 1.4cm,width=0.33\linewidth]{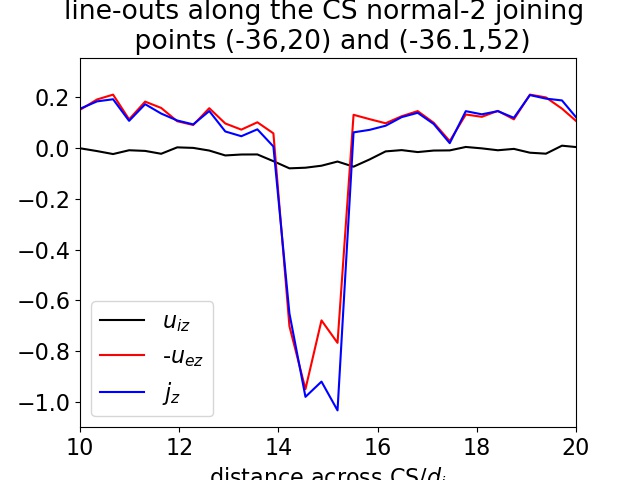}
  \put(-35,15){\large (e)}
  \includegraphics[clip=true,trim=0.0cm 0.5cm 0cm 1.4cm,width=0.33\linewidth]{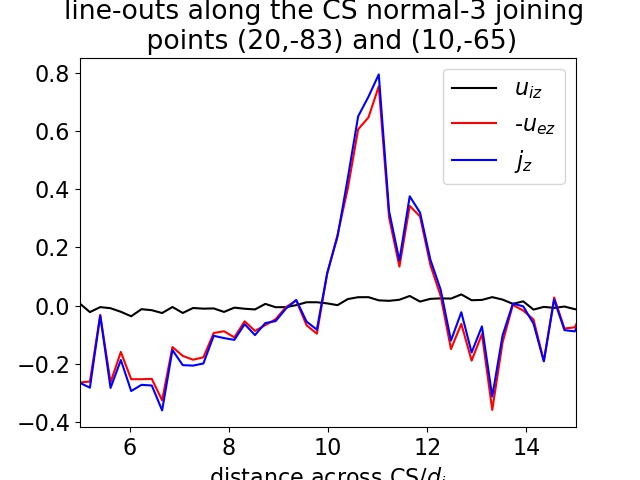}
  \put(-35,15){\large (f)}\\
      \caption{Line-outs of  -$u_{ez}/v_{Ai}$, $u_{iz}/v_{Ai}$ and $J_z/n_0ev_{Ai}$ along the current sheet normals numbered 1 (left column, CS1), 2 (middle column, CS2) and 3 (right column, CS3) in Fig. \ref{fig:jz_t50_t150} at $\omega_{ci}t=50$ for grid resolutions $0.25 d_i \times 0.25 d_i$ (top row) and $0.125 d_i \times 0.125 d_i$ (bottom row).  \label{fig:lineouts_across_CS_t50_high_resolution}}
      \end{center}
\end{figure*}

 We checked the robustness of our results by carrying out simulations with higher grid resolutions and found that our results are not changed. For an example, we show in Fig. \ref{fig:lineouts_across_CS_t50_high_resolution} the line-outs of $-u_{ez}$ , $u_{iz}$ and $J_z$ across the three current sheets for two higher grid resolutions, $0.25 d_i \times 0.25 d_i$ and $0.125 d_i \times 0.125 d_i$. It is clear that current sheets are increasingly due to the electron shear flow as the grid spacing decreases. Other physical quantities (perpendicular bulk velocities, parallel vorticities of electrons and ions and ion temperature anisotropy; not shown here) for higher resolution simulations also show the same behavior as shown in Fig. \ref{fig:lineouts_across_CS_t50}.

\section{Theoretical estimates \label{sec:theory}}
 At the ion kinetic scales, magnetic field frozen into electron bulk velocity is pushed around in the course of turbulence dynamics. The resulting time dependence of  the perpendicular magnetic field $\mathbf{B}_{\perp}$ generates an inductive electric field $E_z$ parallel to the applied magnetic field according to Faraday's law $\nabla \times E_z\hat{z}=-\partial \mathbf{B}_{\perp}/\partial t$.
 Ions are accelerated in the z-direction  by this electric field.  For $\beta_i \sim 1$, ions can be approximated as un-magnetized at the scale of current sheet thickness $\sim  d_i=\rho_i/\sqrt{\beta_i}$. Then the parallel ion bulk velocity evolves as,
 \begin{eqnarray}
   \frac{\partial u_{iz}}{\partial t}=\frac{e}{m_i}E_z,
   \label{eq:duizdt}
 \end{eqnarray}
 where convective derivative $\mathbf{u}_i.\nabla u_{iz}$ has been neglected  in comparison to the time derivative term for the reason  $|(\mathbf{u}_{i}.\nabla)u_{iz}|/|\partial u_{iz}/\partial t| \sim u_{i,\perp}/v_{Ai} \sim 0.1 << 1$ inside current sheets (see Fig. \ref{fig:lineouts_across_CS_t50}d-\ref{fig:lineouts_across_CS_t50}f). Here we have taken $\nabla \sim d_i^{-1}$ and $\partial/\partial t \sim \omega_{ci}$.
 Parallel electron bulk velocity adjusts to satisfy Ampere's law and evolves as,
 \begin{eqnarray}
   \frac{\partial u_{ez}}{\partial t}=\frac{e}{m_i}(E_z-d_i^2\nabla^2E_z),
   \label{eq:duezdt}
 \end{eqnarray}
 obtained by taking time derivative of Ampere's law, neglecting time derivative of plasma density  in comparison to the time derivative of the electron bulk velocity (under the approximation $u_{i,\perp}/v_{Ai} << 1$), and making use of Eq. (\ref{eq:duizdt}) and Faraday's law. Note that Eq. (\ref{eq:duezdt}) is not the same as the electron momentum equation where the time-derivative of electron bulk velocity appears as an electron inertial term, i.e., multiplied by electron mass. In the hybrid simulation model used in this paper, the electron inertial terms in the electron momentum equation are neglected.      

  Estimating $|u_{iz}|\sim |\tau e E_z/m_i|$ and $|u_{ez}|\sim |\tau e E_z (1-d_i^2/L^2)/m_i|$ from Eqs. (\ref{eq:duizdt}) and (\ref{eq:duezdt}), respectively, and $|E_z| \sim L |\mathbf{B}_{\perp}|/\tau$ from Faraday's law,
 \begin{eqnarray}
   \frac{|u_{iz}|}{v_{Ai}} \sim \frac{L}{d_i}\, \frac{|\mathbf{B}_{\perp}|}{B_0}
 \end{eqnarray}
 and,
 \begin{eqnarray}
   \frac{|u_{ez}|}{v_{Ai}} \sim \frac{L}{d_i}\, \frac{|\mathbf{B}_{\perp}|}{B_0}\left|1-\frac{d_i^2}{L^2}\right|.
 \end{eqnarray}
Here $L$ is the scale length of the $E_z$-variation (typically the same as the current sheet thickness) and $\tau$ is the time available for the formation of current sheets before they are disrupted by either the instabilities in current sheets or the turbulence dynamics. Here, we refer to $L$ as the current sheet thickness. The ratio of $|u_{ez}|$ and $|u_{iz}|$ gives,
 \begin{eqnarray}
   \frac{|u_{ez}|}{|u_{iz}|}\sim \left|1-\frac{d_i^2}{L^2}\right|\label{eq:uezDuiz}.
   \end{eqnarray}

   In the limit $L << d_i$ , $|u_{ez}| \propto d_i/L$ and  Eq. (\ref{eq:uezDuiz}) gives $|u_{ez}|/|u_iz| \sim d_i^2/L^2 >> 1$. The thinner the current sheet is, larger (smaller) the parallel electron (ion) bulk velocity and the current in the sheet is increasingly carried by electrons. In current sheets with sub-$d_i$ scale lengths, say $L=0.5\,d_i$, $|u_{ez}|/|u_{iz}| \sim 3$, consistent with the simulation results. 

   The perpendicular electron bulk velocity is simply $|E_{\perp}|/B$, the $E \times B$ drift as per Ohm's law in hybrid plasma models without electron inertia. Outside current sheets, ions are  magnetized and also  execute $E\times B$ drift in the plane of simulation resulting in $|\mathbf{u}_{e\perp}|\approx |\mathbf{u}_{i\perp}|$. They are, however, un-magnetized inside or near current sheets and therefore their in-plane motion deviates from the in-plane electron motion. The difference between perpendicular electron and ion bulk velocities given by Ampere's law, $|\mathbf{u}_{i\perp}-\mathbf{u}_{e\perp}|/v_{Ai}=d_i|\nabla_{\perp}\times B_z\hat{z}|/B_0 \sim (d_i/L) (B_z/B_0)$, is inversely proportional to the gradient scale length resulting in  $|\mathbf{u}_{e\perp}|\neq |\mathbf{u}_{i\perp}|$ in current sheets, as can be seen from Fig. \ref{fig:lineouts_across_CS_t50}d-\ref{fig:lineouts_across_CS_t50}f.
   
   Due to the difference between the perpendicular electron and ion bulk velocities, parallel vorticity also differ in current sheets. This difference can be obtained by taking curl of Ampere's law which gives,
   \begin{eqnarray}
     [\nabla\times\mathbf{u}_i-\nabla\times\mathbf{u}_e]_z=-\omega_{ci}d_i^2\nabla^2B_z
     \end{eqnarray}
The difference is maximum where second derivative of $B_z$ is largest.

\section{Discussion and conclusion  \label{sec:conclusion}}
We carried out two-dimensional PIC-hybrid-code simulations of a
collisionless turbulent plasma ($\beta_i=\beta_e=0.5$) in an external 
magnetic field perpendicular to the simulations plane in a large simulation 
box ($256 d_i \times 256 d_i$). 
We initiated the plasma turbulence by long-wavelength random-phased magnetic 
fluctuations. The simulations show formation of current sheets in turbulent plasma.
We examined the potential sources of free energy available for an 
unstable decay of these current sheets: 
spatial gradients of the plasma density and electron/ion bulk velocities.
By simulations and analytical estimates we could show that the
magnetic-field-aligned (parallel) electron bulk velocity 
$u_{ez} >> |\mathbf{u}_{e\perp}| \sim |\mathbf{u}_{i\perp}|$ and $u_{ez} >> u_{iz}$
dominates the current density $\mathbf{J}=n\,e\,(\mathbf{u}_{i\perp}+{u}_{iz}\hat{z}-\mathbf{u}_{e\perp}-u_{ez}\hat{z})$ 
in ion-scale current sheets, i.e. the sheet current flows primarily parallel to the external
magnetic field. 
The electron bulk velocity component in the direction perpendicular to the magnetic field 
($\mathbf{u}_{e\perp}$), though of the same order of magnitude as the perpendicular ion bulk 
velocity ($\mathbf{u}_{i\perp}$), varies faster than $\mathbf{u}_{i\perp}$ through the
current sheets. 
As a consequence the parallel electron vorticity exceeds and varies faster  through the
current sheets than parallel ion vorticity. 
At the same time the variation of the plasma number density  through the
current sheets is small.

We found that the (half-)thickness of the current sheets formed in the 
turbulent plasma were at most ion inertial lengths $d_i$  in simulations with a 
spatial resolution of 0.5$d_i$.
Simulations with increased grid resolution have shown that  current sheets always 
thin down to the scale of the grid resolution while the peak value of the 
magnetic-field aligned (parallel) current density increases in agreement 
with Eq. (\ref{eq:uezDuiz}).  
 It implies that  current sheets formed in collisionless plasma turbulence have tendencies to thin down below the ion inertial length. In our simulations, thinning of current sheets is stopped by numerical effects at the grid scale. The thinning in real collisionless plasmas, on the other hand, would be stopped by some physical effect at scales below ion inertial length, e.g., at the scale of electron inertial length and/or electron gyro-radius. The thinning might also be stopped by three dimensional plasma instabilities with a wave-vector component along the external magnetic field. Our simulation studies are limited to two-dimensions and ion scales. Further thinning  of current sheets, therefore, cannot be studied by just continuing to decrease grid spacing in our hybrid simulations. Instead, three dimensional plasma models with electron scale physics, at least with electron inertial effects, are necessary to study the thinning of current sheets and free energy sources developed in them at their final thicknesses. With the limitations of  our studies we conclude that electron shear flow develops in ion scale current sheets during their formation in collisionless plasma turbulence.

  Kinetic scale current sheets with current mainly due to the electron bulk velocity, i.e., with electron shear flow structure, were found recently by MMS spacecraft in Earth's magnetotail \citep{hubbert2021}. Fig. 1 in their paper \citep{hubbert2021} shows a sub-ion scale current sheet with $v_{eM}\sim 1000$ km/s $>> v_{iM} \sim 50$ km/s in the region where $B_L$ reverses its sign. Here $v_{e}$ and $v_i$ are electron and ion bulk velocities, respectively, and the suffix (L, M, N) represent the local current sheet coordinate system (``N'' represents the direction of current sheet normal, ``M'' the direction of the main current and ``L'' the direction orthogonal to both M- and N-directions).  The relative variation of $v_{eM}$ (from $\sim$ 100 km/s at the current sheet edge to $\sim$ 1000 km/s at the current sheet center) is much larger than the relative variation of electron/ion density (for electrons from 0.6 cm$^{-3}$ at current sheet edge to 0.7 cm$^{-3}$ at the current sheet center, less than 20\%) in the field reversal region.
  Development of automated methods for the detection of current sheets in space observations~\citep{khabarova2021} and numerical simulations~\citep{azizabadi2021} of plasma turbulence is required to study the detailed structure of current sheets formed in plasma turbulence. Studies on current sheet equilibria based on such detection  would be instrumental to understand the role of plasma instabilities in collisionless dissipation~\citep{zelenyi2020}.

Note that in laboratory reconnection experiments with strong external (guide-) 
magnetic fields like the Greifswald VINETA.II experiment sheets of
electron dominated current flows parallel to the external magnetic field and
electron shear flows in the perpendicular direction were found~\citep{stechow2016}. 
Hence, the formation of such thin current sheets seems to be generic in 
collsionless plasmas with external magnetic field.  
Electron-magnetohydrodynamic simulations carried out to understand the results of the laboratory experiments revealed that plasma instabilities driven 
by perpendicular electron shear flows generate magnetic fluctuation as they
were found in the experiments, if the  electron inertia is taken into 
account~\citep{jain2017a}.


\begin{acknowledgments}
We gratefully acknowledge the developers of the A.I.K.E.F. code and the financial 
support by the German Science Foundation (DFG), project JA 2680-2-1.

Part of the simulations were carried out on the HPC-Cluster of the Institute for 
Mathematics at the TU Berlin and the Max-Planck-Institute for Solar System 
Research G\"ottingen.
.

\end{acknowledgments}


\bibliography{references_2DIKPT}{}
\bibliographystyle{aasjournal}



\end{document}